\documentclass[fleqn,usenatbib]{mnras}

\usepackage{newtxtext,newtxmath}

\usepackage[T1]{fontenc}
\usepackage{subfigure}

\DeclareRobustCommand{\VAN}[3]{#2}
\let\VANthebibliography\thebibliography
\def\thebibliography{\DeclareRobustCommand{\VAN}[3]{##3}\VANthebibliography}

\usepackage{graphicx}	

\usepackage{amsmath}	
\usepackage{amssymb}	
\usepackage{multicol}   
\usepackage{bm}
\usepackage{color}
\usepackage{xspace}
\usepackage{CJKutf8}

\definecolor{mygray}{gray}{0.6}
\definecolor{TsinghuaPurple}{cmyk}{0.58,0.90,0,0}
\definecolor{magenta}{rgb}{0.858, 0.188, 0.478}
\definecolor{orange}{rgb}{1.0, 0.4, 0.0}

\newcommand{\ccc}[1]{\textcolor{orange}{[\textit{\small #1}]}}
\newcommand{\cim}[1]{\textcolor{orange}{[\textbf{$\star$\small #1$\star$}]}}
\newcommand{\hjc}[1]{\textcolor[RGB]{0,155,155}{[\textit{\small #1}]}}
\newcommand{\hjadd}[1]{\textcolor[RGB]{0,155,155}{#1}}

\newcommand{\hjch}[2]{\textcolor{mygray}{\sout{#1}}{\textcolor[RGB]{0,155,155}{{#2}}}}
\newcommand{\corem}[1]{\textcolor{mygray}{\sout{#1}}}
\newcommand{\hjrem}[1]{\textcolor{mygray}{\sout{#1}}}
\newcommand{\xxx}[1]{\textcolor{blue}{\textbf{xxx}\xspace}}

\newcommand{\fg}[1]{Fig.~\ref{fig:#1}}
\newcommand{\Fg}[1]{Figure~\ref{fig:#1}}

\newcommand{\eq}[1]{Eq.~(\ref{eq:#1})\xspace}
\newcommand{\Eq}[1]{Equation~(\ref{eq:#1})\xspace}

\newcommand{\tb}[1]{Table~\ref{tab:#1}\xspace}
\newcommand{\se}[1]{Sect.~\ref{sec:#1}\xspace}
\newcommand{\Se}[1]{Section~\ref{sec:#1}\xspace}

\newcommand{\sef}[1]{\ref{sec:#1}\xspace}
\newcommand{\App}[1]{Appendix~\ref{app:#1}\xspace}

\usepackage{ulem}
\makeatletter
\def\uwave{\bgroup \markoverwith{\lower3.5\p@\hbox{\sixly \textcolor{red}{\char58}}}\ULon}
\font\sixly=lasy6 
\makeatother

\renewcommand{\cim}[1]{}
\renewcommand{\ccc}[1]{}
\renewcommand{\corem}[1]{}

\renewcommand{\uwave}[1]{#1}

\renewcommand{\hjc}[1]{}
\renewcommand{\hjrem}[1]{}
\renewcommand{\hjadd}[1]{#1}
\renewcommand{\hjch}[2]{#2}

\title[Planet formation in ALMA rings]{Efficient planet formation by pebble accretion in ALMA rings}

\author[Jiang \& Ormel]{
Haochang Jiang (\begin{CJK*}{UTF8}{gbsn}蒋昊昌\end{CJK*})$^{1,2}$\thanks{E-mail: jhc19@mails.tsinghua.edu.cn},
Chris W. Ormel$^{1}$\thanks{E-mail: chrisormel@tsinghua.edu.cn}
\\
$^{1}$Department of Astronomy, Tsinghua University, Haidian DS 100084, Beijing, China  \\
\hjadd{$^{2}$European Southern Observatory, Karl-Schwarzschild-Str 2, 85748 Garching, Germany}
}

\date{Accepted XXX. Received YYY; in original form ZZZ}

\pubyear{2022}

\begin{document}
\label{firstpage}
\pagerange{\pageref{firstpage}--\pageref{lastpage}}
\maketitle

\begin{abstract}
    In the past decade, ALMA observations have revealed that a large fraction of protoplanetary discs contains rings in the dust continuum. These rings are the locations where pebbles accumulate, which is beneficial for planetesimal formation and subsequent planet assembly. We investigate the viability of planet formation inside ALMA rings in which pebbles are trapped by either a Gaussian-shaped pressure bump or by the strong dust backreaction. Planetesimals form at the mid-plane of the ring via streaming instability. By conducting N-body simulations, we study the growth of these planetesimals by collisional mergers and pebble accretion. Thanks to the high concentration of pebbles in the ring, the growth of planetesimals by pebble accretion becomes efficient as soon as they are born. We find that \hjch{type-I}{planet} migration plays a decisive role in the evolution of rings and planets. For discs where planets can migrate inward from the ring, a steady state is reached where the ring spawns ${\sim}20 M_\oplus$ planetary cores as long as rings are fed with materials from the outer disc. The ring acts as a long-lived planet factory and it can explain the “fine-tuned” optical depths of the observed dust rings in the DSHARP large program. In contrast, in the absence of a planet removal mechanism (migration), a single massive planet will form and destroy the ring. A wide and massive planetesimals belt will be left at the location of the planet-forming ring. Planet formation in rings may explain the mature planetary systems observed inside debris discs.
\end{abstract}




\section{Introduction}
Protoplanetary discs are the cradle of planets. In the past two decades, more than 5000 exoplanets have been found, demonstrating a variety in configurations of exoplanet systems. At the same time, thanks to the progress in its high sensitivity and spatial resolution, the Atacama Large Millimeter/submillimeter Array (ALMA) reveals unprecedented images of planet-forming discs in both continuum and line-emission \citep[e.g.,][]{AndrewsEtal2018,LongEtal2018,CiezaEtal2019,OebergEtal2021}. The ALMA continuum observation of protoplanetary discs allows the measurement of the mass budget of (sub)millimeter size particles, a.k.a pebbles, which are the building blocks for planets.

One striking feature of protoplanetary discs is the ubiquitous rings and gaps \hjadd{in large (radius $\gtrsim$50~au) discs \citep[e.g.,][]{HuangEtal2018,LongEtal2019,CiezaEtal2021}}, which challenges the classical framework where discs are considered as ‘smooth’ \citep[e.g.,][]{Weidenschilling1977i,Hayashi1981}. \hjadd{Even though the existing high-resolution observations are generally biased towards brighter and more massive discs \citep[][]{LongEtal2019,Andrews2020}, }\hjrem{A large fraction of the protoplanetary discs observed with ALMA display} near-perfect annular rings in the continuum \hjch{, in which enormous quantities of dust appear to be concentrated,}{are found across stars covering a large range of spectral types and ages} \citep[e.g.,][]{vanderMarelEtal2019}. The formation of these rings is an open question. The most popular hypothesis is that rings are pebbles trapped by a local pressure maxima \citep[e.g.,][]{DullemondEtal2018}. Due to aerodynamical drag, the radial velocity of pebbles in protoplanetary discs follows the gas pressure gradient. A local pressure bump is, therefore, a site for dust trapping. A number of mechanisms can form such pressure bumps in protoplanetary discs, including dead zone boundaries \citep[e.g.,][]{FlockEtal2015}, zonal flows \citep[e.g.,][]{BaiStone2014}, photoevaporation \citep[e.g.,][]{OwenKollmeier2019}, snowlines \citep[e.g.,][]{KretkeLin2007,OkuzumiEtal2016}, and the most widely proposed mechanism --- planets \citep[e.g.,][]{DongEtal2015,ZhangEtal2018}. 

The planet-disc interaction provides a plausible way of explaining the disc structure. In particular, the detection of localized kinematic deviations from Keplerian motion in $^{12}$CO maps in a few discs makes the planet explanation compelling \citep[e.g., HD~163296,][]{TeagueEtal2018,PinteEtal2020}. Yet, the confirmation of direct evidence of planets inside discs is challenging. One promising way is by the H line emission, which originates from an accretion front at the planetary surface. Dozens of discs have been targeted by direct imaging facilities, but PDS~70 remains the only system where accreting exoplanets were unambiguously identified \citep[e.g.,][]{KepplerEtal2018,HaffertEtal2019}. In addition, the MAPS Large program released tens of line-emission maps of five protoplanetary discs, allowing the assessment of the correlation between dust substructures and gas substructures \citep{LawEtal2021}. Yet, statistical tests on the correlation between dust continuum and gas substructures report no significant correlation \citep{JiangEtal2022}. Therefore, it is debatable whether planets, or more generally, pressure bumps, are the primary cause for rings in discs, which questions the frequently made link between giant planets and the rings and gaps.

In addition, forming a planet by core accretion at \hjch{these great distances}{the location of the disc substructures, which are often inferred at large distances (${>}20$~au) \citep[e.g.,][]{HuangEtal2018,LongEtal2019,CiezaEtal2019,CiezaEtal2021},} requires time. Even the acclaimed pebble accretion \citep[][]{OrmelKlahr2010,LambrechtsJohansen2012} for which the accretion rate is amplified by gas drag,
it takes the disc lifetime for a planetary core to grow at wide orbits (${>}10$~au) in \textit{smooth} disc \citep{Ormel2017}. \hjch{On the other hand, from the observational aspect,}{Observationally, however,} an increasing number of rings are found in young class 0/I disc \citep[e.g.,][]{SheehanEisner2017,SheehanEisner2018,deValonEtal2020,Segura-CoxEtal2020,SheehanEtal2020}. The age of these discs are less than $0.5$ Myr, further exacerbating the timescale problem.

Alternatively, giant planets on distant orbits could also form by gravitational instability (GI) in the disc \citep[e.g.,][]{Boss1997}. Yet, a study on the mass function of direct\hjadd{ly}-imaged planets suggests GI might not be the primary source of giant planet formation \citep{WagnerEtal2019}. Moreover, for the potential planets that are assumed to be associated with the ALMA rings and gaps, the inferred mass range covers a large parameter space and strongly depends on the unknown level of disc turbulence. In particular, some potential "planets" are inferred to be only several earth mass \citep{WangEtal2021}, while the planet formed by GI is usually a gas giant. Recently, \citet{DengEtal2021} argues that this problem can be solved if the disc is magnetized, and the planet could be formed via the magnetically controlled disc fragmentation. However, the model relies on the strong magnetic field which is largely unknown in the disc. These possible intermediate-mass planets are overall hard to explain by gravitational instability.

While the origin of the rings is not well understood, due to the local enhancement of the dust-to-gas ratio, rings are believed to be perfect sites of planetesimal formation by streaming instability and self-gravity collapse of pebbles \citep[e.g.,][]{CarreraEtal2015,YangEtal2017,LiYoudin2021,CarreraEtal2021,KlahrSchreiber2021,CarreraEtal2022,XuBai2022i}. Tens of Earth mass in solids are accumulated within these rings, which may allow planetesimal formation up to the same magnitude \citep[e.g.,][]{StammlerEtal2019,JiangOrmel2021}. Whether or not these planetesimals can grow into planets is an intriguing question, similar to the age-old chicken or egg dilemma---are rings in discs caused by planets that were already formed, or are rings intrinsic properties of discs where planets form?

There are two interpretations of how pebble rings will promote the formation and growth of planets. One group of studies focuses on the evolution of a swarm of planetesimals formed from a pebble ring in smooth discs. By assuming planetesimals formed from streaming instability, \citet{LiuEtal2019} study the growth of a planetesimal belt formed from a ring at the ice line. A super-Earth planet can quickly form by a hybrid approach of planetesimal accretion and pebble accretion. \citet{JangEtal2022} extend the study by placing the ring at a number of different radii. However, because of the lower pebble surface density, higher aspect ratio, and longer orbital timescale at the outer disc, the formed planetesimals can hardly grow at wide orbital distances (${\gtrsim}30$~au), which is the location of the ALMA rings. Yet in both these works the \hjch{aerodynamics of pebbles is treated in a smooth}{radial} pressure gradient \hjch{where}{stays negative and} pebbles always efficiently drift inward. The enhancement of the pebble density at the ring location is not taken into account.

Alternatively, another group of studies focuses on the pebble aerodynamics inside a pressure bump. As a pilot study on this question, \citet{Morbidelli2020} analytically solves the growth of a single Mars-mass core inside the B77 ring in Elias~24 by assuming the ring is supported by a permanent pressure bump. They find the embryos grow very slowly at 75~au (which can only grow to $1 M_\oplus$ after $\sim3$ Myr). On the other hand, the growth of the embryo turns very efficient when putting the same ring at 5 au, where the embryos can grow to $10 M_\oplus$ within $0.5$ Myr. Even faster growth is reported in \citet{GuileraEtal2020}, where a $13 M_\oplus$ planet can form in only $0.02$~Myr at a bump located at the water snowline (3 au). Recently, \citet{Chambers2021} and \citet{LeeEtal2022} revisit this question, and both of them find the growth of embryos can become much more efficient when taking larger St and/or a more massive ring.  Yet, in these works, only one isolated embryo is considered, and the output is sensitive to the initial mass of this embryo. However, realistically, a large number of planetesimals will form as demonstrated in \citet{LiuEtal2019} and \citet{JangEtal2022}. In addition, the mutual interaction of planetesimals may influence the evolution significantly \citep[e.g.][]{LevisonEtal2012}.

In this paper, we study the planetesimal formation and the following-up growth of the planetesimal via pebble accretion inside a pebble ring consistently. Two different assumptions on the ring nature are considered-- the clumpy ring and pressure bump supporting ring. The clumpy ring model assumes that the ring is a manifestation of dense, clumpy mid-plane pebbles in a \textit{smooth} gas disc \citep[][hereafter JO21]{JiangOrmel2021}. In this model, due to the strong back-reaction, the clumpy medium itself hardly experiences radial drift. The accumulation of pebbles results in positive feedback to slow down the dust drift, and eventually leads to a traffic jam shown as a pebble ring.

In the clumpy ring model, because of the high dust-to-gas ratios, pebbles get into the clumpy state within the ring and their drift is significantly decelerated. A similar condition is met in the ring induced by a pressure bump. Therefore, the pebble accretion threshold will be much lower inside the ring. Thus, in the interval where clump pebbles and planetesimals co-exist, (small) planetesimals could efficiently feed from pebbles. On the other hand, mutual scattering of planetesimals would oppose the growth \citep[e.g.,][]{SchoonenbergEtal2019}. In this work, we investigate the viability of planet formation in ALMA rings by conducting N-body simulations that account for planetesimal dynamics, pebble accretion, and planet migration. We also contrast the standard pressure bump setup with the clumpy ring interpretation proposed by JO21.

The plan of the paper is as follows. We describe the theoretical background and numerical setup in \Se{model}. We present and explain the numerical results in \Se{results}, which are further discussed in \Se{discussion}. We summarize the main results and conclusions in \Se{conclusions}.

\section{Model}\label{sec:model}

\begin{figure*}
    \centering
    \includegraphics[width=0.99\textwidth]{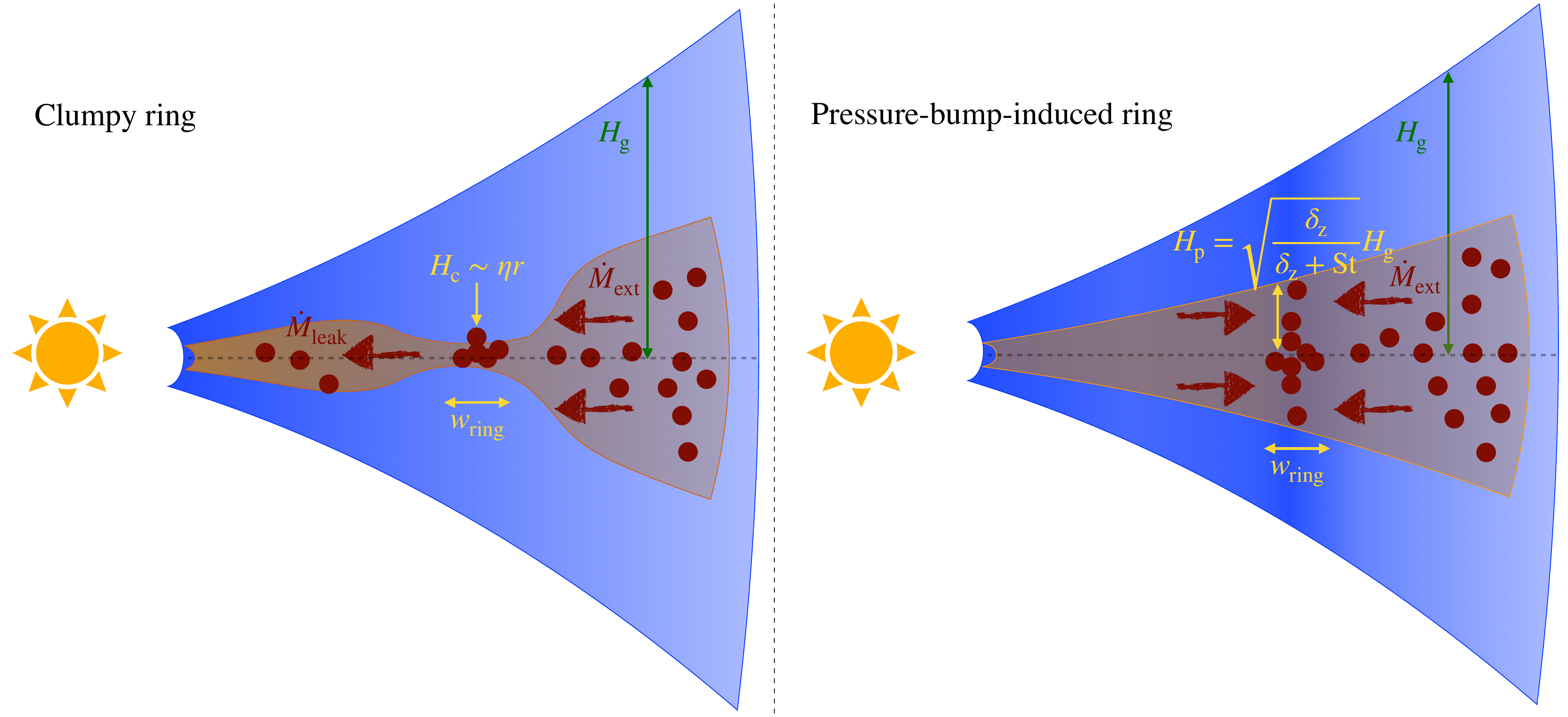}
    \caption{\label{fig:CRvsPB} Sketch illustrating the differences between the clumpy ring model (left) and the ring induced by a pressure bump (right). The clumpy ring is where pebbles experience strong pile-ups because the dust backreaction slows down the radial drift of pebbles. The pebble scaleheight is determined by the Kelvin-Helmholtz instability at the ring location in the CR model, while the pebbles are still well coupled with gas in PB-induced rings. Pebbles can still leak from the clumpy ring to the inner disc ($\dot{M}_{\rm leak}$). In contrast, we assume that the pressure bump will fully filter the pebbles from the inner disc.}
\end{figure*}

We test two different types of planetesimal-forming rings in this work where we group them by distinguishing the nature of the pebble concentration inside the ring:
\begin{enumerate}
    \item Ring induced by a permanent pressure bump (PB)
    \item Ring sustained by aerodynamical effects only, in the context of the clumpy ring (CR) model.
\end{enumerate}
\Fg{CRvsPB} shows a cartoon that summarizes the differences between CR and PB models. For the PB model, \citet{Morbidelli2020} analytically solves for the planet growth by pebble accretion for the B77 ring in Elias~24. The setup of the dust ring is similar in our work with the one used by them. However, they directly initialized a Mars mass embryo inside their framework and did not account for the preceding planetesimals coagulation process. Instead, we take into account the formation of planetesimal self-consistently and numerically solve the problem in N-body simulations, which produces a more realistic outcome. The planetesimal formation approach is based on the assumption that the dust concentration may induce streaming instability, which can fulfill the criteria for forming planetesimals. The details of planetesimal formation will be discussed in \se{pltsfm}.

The other scenario we studied is that of the clumpy ring model. In the CR model, it is assumed that ALMA rings are the locations where pebbles have collapsed as their number is too great for turbulence to sustain \citep{YoudinShu2002}. Still, as they are not contained by pressure, pebbles "leak" \hjch{at}{from} the ring \hjch{front}{inner} edge, requiring a continuous supply of pebbles from the exterior disc. \hjrem{An intriguing feature of the clumpy ring model} In JO21 \hjadd{it was observed that rings could "migrate" outward, at rates of ${\sim}10\,\mathrm{au\,Myr}^{-1}$, as pebbles piled up at its outer edge.} \hjrem{"outward migration" of the ring, or more accurately, of the shape of the ring.} \hjch{This is due to the fact that external mass flux coming from the outer disc and the new traffic jam always happens at the tail of the queue and the head of the queue will be leaked the first. In JO21 where the rings are formed}{And the ring profile was obtained} by solving the transport equations \hjadd{for pebbles}. \hjadd{But In this work, in order to facilitate a} direct comparison \hjadd{between} the PB and CR models, we consider \hjch{a scenario where the ring features no migration}{for simplicity that the ring is stationary and long lived}. 
Such an assumption applies when the mass flux feeding the ring and the mass flux \hjch{losing from}{leaving} the ring reach a perfect balance. \hjrem{This scenario is also frequently found in JO21 when the radial transport of clumps is moderate rather than totally halted, and therefore the outward migration of the shape of the ring will be canceled with the advection of clumpy pebbles}.


\subsection{Disc structure}
In our study, we focus on the rings observed by ALMA \citep[e.g.,][]{AndrewsEtal2018,LongEtal2018}. We take the B74 ring in AS~209 as an example, whose ring location is $r_0 = 74.2\,\rm au$, ring width is $w_{\rm ring} = 3.38\,\rm au$ and the disc mid-plane temperature is
$T_0=16\,\rm K$ at the ring location \citep{DullemondEtal2018}. For simplicity, we choose the stellar mass $M_\star = 1 M_\odot$. The surface density and temperature of the gas in the smooth disc is given by
\begin{equation}
    \Sigma_{\rm g} = \Sigma_{\rm g,0} \times \left(\frac{r}{r_0}\right)^{-1} 
\end{equation}
\begin{equation}\label{eq:Td}
    T = T_0\times\left(\frac{r}{r_0}\right)^{-0.5}
\end{equation}
where $\Sigma_{\rm g,0} = 3.6\,\mathrm{g}\,\mathrm{cm}^{-2}$. We assume that the disc is isothermal in the vertical direction. The isothermal sound speed is $c_s = \sqrt{k_B T/\mu m_{\rm H}}$, with $m_{\rm H}$ the proton mass and $\mu = 2.3$ the mean molecular weight in atomic units. The disc is flared, and the aspect ratio is
\begin{equation}\label{eq:cs}
    h_{\rm g} \equiv \frac{H_\mathrm{g}}{r} \equiv \frac{c_s}{\Omega_K r} = h_0\times\left(\frac{r}{r_0}\right)^{0.25} 
\end{equation}
where $\Omega_K(r) = \sqrt{GM_\star/r^3}$ is the local Keplerian frequency and $H_{\rm g}$ is the gas scaleheight. With substitutions, the aspect ratio is $h_0 = 0.07$ at the characteristic location $r_0$. These parameters we choose are consistent with the classical $\alpha$-accretion disc model \citep{ShakuraSunyaev1973} where the gas mass accretion rate $\dot{M}_g = {3\pi \alpha c_s H_{\rm g}\Sigma_{\rm g}} \equiv 10^{-9} \times (\alpha/10^{-3}) M_\odot \rm yr^{-1}$, where $\alpha$ is the coefficient of gas viscosity.

\subsubsection{Pressure bump induced ring (PB)}
For the case where the pebble ring is formed by a prominent pressure bump, we directly follow the calculation in \citet{DullemondEtal2018} and \citet{Morbidelli2020} to set up the pebble ring. The key assumptions is that the radial pressure profile follow a nominal Gaussian shape around the ring:
\begin{equation}\label{eq:pr} 
    p(r) = p_0 \exp\left(-\frac{(r-r_0)^2}{2w_{\rm pb}^2}\right)
\end{equation}
where $r_0$ and $w_{\rm pb}$ are respectively the central location and width of the pressure bump. Since the radial pressure support, the azimuthal gas velocity $v_\phi$ is slightly different from the Keplerian velocity $v_K=\Omega_K r$. The dimensionless measure of the radial pressure gradient is \citep{NakagawaEtal1986}
\begin{equation}\label{eq:eta}
    \eta \equiv = \frac{v_K - v_\phi}{v_K} = -\frac{1}{2}\frac{c_s^2}{v_K^2}\frac{\partial \log{P}}{\partial \log{r}}.
\end{equation}
For the case with a pressure bump, by substituting \eq{pr}, we obtain
\begin{equation}\label{eq:eta_pb}
    \eta_{\rm pb} = \frac{1}{2} \frac{r(r-r_0)}{w_{\rm pb}^2} h_{\rm g}^2.
\end{equation}
Due to gas drag, the radial velocity of pebbles follows
\citep{Weidenschilling1977}
\begin{equation}\label{eq:vdr}
    v_\mathrm{dr} = -\frac{2\mathrm{St}}{1+\mathrm{St}^2} \eta v_K
\end{equation}
where St is the pebble's Stokes number. The pebble will concentrate towards the pressure bump centre. Yet, the pebble also experiences radial diffusion with corresponding diffusivity \citep{YoudinLithwick2007}
\begin{equation}
    D_\mathrm{d,r} = \frac{\delta_{\rm r} H_{\rm g}^2\Omega_K}{1+{\rm St}^2}
\end{equation}
where we assume $\delta_{\rm r} = \alpha$ as a usual choice \citep[e.g.,][]{PinillaEtal2021}.
By balancing radial drift from gas drag with dust diffusion, the surface density profile of the pebble ring reads
\begin{equation}\label{eq:Sigr}
    \Sigma_{\rm r}(r) = \Sigma_{\rm r,0} \exp\left(-\frac{(r-r_0)^2}{2w_{\rm ring}^2}\right)
\end{equation}
where $\Sigma_{\rm r,0}$ is the peak surface density of the ring. The ring width $w_{\rm ring}$ scales with the width of the pressure bump
\begin{equation}\label{eq:wring}
    w_{\rm ring} = w_{\rm pb}\sqrt{\frac{\delta_{\rm r}}{\delta_{\rm r} + \rm St}}
\end{equation}
and the pebble mass of the ring can be calculated by
\begin{equation}\label{eq:M_ring}
    M_{\rm ring} = (2\pi)^{1.5}r_0 w_{\rm ring}\Sigma_{\rm r,0}
\end{equation}
As argued by \citet{Morbidelli2020}, since the separation of the dust rings observed in ALMA is typically much larger than the ring width, these formulas should hold at least at the range of several $w_{\rm wring}$ away from the ring centre. In the background disc away from the ring, for simplicity, we do not solve for the pebble density distribution, but use a constant pebble flux, see \se{PA} for detailed treatment. However, pebble accretion from the pressure-supported background disc occurs at low efficiency at distant orbits. Pebble accretion outside the ring region is significantly smaller in most of our simulations compared with pebbles accreted from the ring.

Finally, for the PB model, we use the $\alpha$-prescription with dimensionless diffusivity parameter $\delta_{\rm z}$ for vertical transport of the pebbles. Assuming a vertical Gaussian distribution for the volume densities of both pebble and gas, the scaleheight of the pebble can be obtained by balancing the dust settling with vertical diffusion \citep{DubrulleEtal1995,BirnstielEtal2010} 
\begin{equation}\label{eq:Hd}
    H_{\rm p} = \sqrt{\frac{\delta_\mathrm{z}}{\delta_\mathrm{z}+\mathrm{St}}}H_\mathrm{g}
\end{equation}

For most of our simulations, we set the values of the gas viscosity $\alpha$ and pebbles' diffusivity $\delta_{\rm r}$, $\delta_{\rm z}$ to be $10^{-3}$, which is suggested by observation of dust distribution geometry in several discs and widely used in theoretical studies \citep[e.g.,][]{PinteEtal2016,DullemondEtal2018,VillenaveEtal2020}.
Yet, values of these parameters remain uncertain \citep[see][and references therein]{PinillaEtal2021,MiotelloEtal2022}.
For example, depending on the ring width and CO kinematics of HD~163296 and AS~209, \citet{RosottiEtal2020} suggest $\delta_{\rm r} = 10^{-2}$  by assuming the balance between gas diffusion and dust decoupling in the radial direction, while \citet{VillenaveEtal2022} finds $\delta_{\rm z} = 10^{-5}$ based on the thin dust layer of the edge-on disc in Oph~163131. 
Moreover, it is also suggested that $\delta_{\rm r}$ and $\delta_{\rm z}$ can be different in simulation with non-ideal MHD effects \citep[e.g.,][]{BaiStone2014,XuEtal2017,BaehrZhu2021}. Without loss of generality, we also test the impact of lower disc turbulence ($\delta_{\rm z} = 10^{-4}$) on our model in parameter studies.

\subsubsection{The clumpy ring (CR)}
As explained at the start of this section, we assume a stationary ring for the CR model in order to facilitate a direct comparison with the PB model. In addition, we approximate the CR profile as Gaussian, \eq{Sigr}, with the same width $w_{\rm ring} = 3.38\,$au.

In the clumpy ring model, these clumpy pebbles are concentrated in a thin mid-plane of the disc. Following \citet{Sekiya1998} and \citet{YoudinShu2002}, JO21 assumes that clump pebbles reside in the mid-plane with the scaleheight set by Kelvin--Helmholtz instability
\begin{equation}\label{eq:Hc}
    H_{\rm c} =\sqrt{Ri_{\rm c}} \eta r \Psi(\psi)
\end{equation}
Hereafter referred to as the clump layer. For the critical Richardson number $Ri_{\rm c}$, \citet{Chiang2008} points out that when the vertically-integrated dust-to-gas ratio is between 1 and 5 times solar abundance, the Richardson number $Ri_{\rm c} \sim 0.1$. Thus we take $Ri_{\rm c} = 0.1$ in this paper. The dimensionless self-gravitational term $\Psi(\psi) \equiv \sqrt{1+2\psi} - \psi \ln{[(1+\psi+\sqrt{1+2\psi})/\psi]}$ is order of unity, where
\begin{equation}
    \psi \equiv \frac{4\pi G \rho_\mathrm{g}}{\Omega_K^2} = 0.16 \times  
    \left(\frac{\dot{M}_g}{10^{-9}M_\odot\,\rm yr^{-1}}\right)
    \left(\frac{\alpha}{10^{-3}}\right)^{-1}
    \left(\frac{M_\star}{M_\odot}\right)^{-1.5}
    \left(\frac{r}{r_0}\right)^{0.75}
\end{equation}
is the dimensionless self-gravity parameter of gas.


\subsubsection{Pebble properties}
\hjadd{For simplicity, we work with a constant Stokes number in every simulation. This would match a scenario in which the growth of pebbles is limited by a fragment barrier \citep[e.g.][]{BirnstielEtal2011}
\begin{equation}
    {\rm St_{frag}} = \frac{v_{\rm frag}^2}{3 \delta_{\rm t}c_s^2} 
\end{equation}
where $v_{\rm frag}$ is the fragmentation velocity, $\sqrt{\delta_{\rm t}}c_s$ is the root-mean-square turbulent velocity. Motivated by the low sticking velocity found in experiment for CO$_2$ ice \citep{MusiolikEtal2016}, cold and dry H$_2$O ice \citep{GundlachEtal2018,MusiolikWurm2019}, and/or high-porosity ice \citep{SchraeplerEtal2022}, we take $v_{\rm frag} = 2 \rm m\,s^{-1}$ as the default value. With $\delta_{\rm t} = 10^{-3}$, this leads to $ {\rm St_{frag}} = 0.01$, corresponding to pebble radius of $200\,\mu\mathrm{m}$ size at $r_0$. Such a value is also consistent with the millimeter polarization measurements for several protoplanetary discs \citep[e.g.,][]{KataokaEtal2015,YangEtal2016,LinEtal2020} and some results of multiple wavelength analysis \citep[e.g.,][]{MaciasEtal2021,SierraEtal2021,GuidiEtal2022} in ALMA observation. We therefore opt for $\mathrm{St}=0.01$ as our default value, but we will investigate how the Stokes number affects our results.}

\subsection{Planetesimal formation}\label{sec:pltsfm}
We focus on the hypothesis that planetesimal are formed via streaming instability (hereafter SI) in this work. Planetesimals formation is found to be natural output of SI in hydrodynamical simulations \citep[e.g.,][]{CarreraEtal2015,CarreraEtal2021,YangEtal2017,LiEtal2019}. A widely used criterion for planetesimal formation in disc is the volume dust-to-gas ratio exceeds unity \citep[e.g.,][]{JohansenEtal2006}. \hjch{Very recently, with improved numerical resolution, Li \& Youdin (2021) found this threshold can be slightly lower for pebbles with $\rm St>0.01$, which make the planetesimal formation more efficient. However, this does}{However, recent SI simulations show that the criterion of starting SI is more complex than the simple mid-plane dust-to-gas ratio $Z \equiv \rho_{\rm d}/\rho_{\rm g} = 1$ \citep[e.g.,][]{CarreraEtal2015,YangEtal2017,LiYoudin2021}. Specifically, with improved numerical resolution, \citet[][]{LiYoudin2021} found the typical required dust-to-gas is only $0.5$ for pebbles exceeding ${\rm St}>0.01$. Conversely, for pebbles of Stokes number less than ${\rm St}<0.01$, they find that $Z>2$ is needed for SI \citep[see ][for more discussion]{LiYoudin2021}. 
In addition, if a particle size distribution is present, streaming instability may not occur so efficiently as in the single size pebble simulations \citep[e.g.,][]{KrappEtal2019}\footnote{However, \citet[][]{ZhuYang2021} find that the multiple-pebble-size simulation could be divided into two distinct regimes: fast or slow growth. When the dust-to-gas ratio exceeds unity, as defined in our condition, it always falls into the fast-growth regime, where the the saturation states is similar to the single-species counterpart \citep[][]{YangZhu2021}.}. Overall, all of the effects above do} not qualitatively affect the picture of planetesimal formation via SI and in our model we take for simplicity the criterion as $Z=1$ independent of St.  

We integrate the total mass located inside the planetesimal forming region (where $\rho_\mathrm{d}>\rho_\mathrm{g}$ at the disc mid-plane), and the planetesimal formation rate is \citep[e.g.,][]{DrazkowskaEtal2016,SchoonenbergOrmel2017}
\begin{equation}\label{eq:dotM_plt}
    \dot{M}_{\rm plt} = \zeta\frac{M(\rho_\mathrm{d}>\rho_\mathrm{g})}{t_{\rm sett}}
\end{equation}
where $\zeta$ is the planetesimal formation efficiency and $t_{\rm sett} = 1/{\rm St}\Omega_K$ is the settling timescale. 

In our N-body simulation, we assume that all planetesimals are formed with the same initial mass $m_{\rm 0}$. For every $1000\,\rm yr$, we calculate the planetesimal formation rate and the total number of planetesimal formed in the past $1000\,\rm yr$. We add these planetesimals randomly inside the forming region. We ensure that each newly formed planetesimal is separated from other planetesimals by at least 5~Hill radii. We set the initial eccentricity $e_0 = 10^{-5}$ and the initial inclination to be $i_0=e_0/2$. We tested smaller values of $e_0$ and $i_0$ and find that since the orbits of planetesimals are readily excited, the results are insensitive to the initial values.

According to SI simulations \citep[e.g.,][]{SchaeferEtal2017}, the mass of the planetesimal formed by SI shows a top-heavy initial mass function with a cutoff at the massive end. However, as demonstrated by \citet{LiuEtal2019}, the most massive planetesimal would dominate the following mass growth and dynamical evolution of the whole population. We choose the planetesimal masses accordingly. By reviewing the literature on SI studies \citep{JohansenEtal2015,SimonEtal2016,SchaeferEtal2017,AbodEtal2019,LiEtal2019}, \citet[][see their Table 1]{LiuEtal2020} give a fit for the initial mass of the planetesimal
\begin{equation}\label{eq:m0}
\begin{aligned}
    \frac{m_{\rm 0}}{M_\oplus} =& 2\times10^{-3}
    \left(\frac{f_{\rm plt}}{40}\right)
    \left(\frac{C}{5\times10^{-5}}\right)
    \left(\frac{\psi}{0.16}\right)^{a+1}
    \left(\frac{h_{\rm g}}{0.07}\right)^{3+b}
    \left(\frac{M_\star}{M_\odot}\right)    \\
\end{aligned}
\end{equation}
where the values of numerical factors ($a=0.5$, $b=0$, $C=5\times10^{-5}$) are chosen to be the same as in \citet[][see their Section 2.4 for fruitful discussion on choice of these parameters]{LiuEtal2020}. We also control the initial planetesimal mass through $f_{\rm plt}$, which is the ratio between the maximum mass and the characteristic mass. The typical value falls between tens to thousands in SI simulation \citep[and $f_\mathrm{plt} = 400$ in][]{LiuEtal2020}. We test initial masses ranging from Ceres to Mars, which corresponds to $f_{\rm plt}=2\,\rm to\,2000$. For the fiducial value, we choose $f_{\rm plt}=40$, which is consistent with a Pluto mass body. Surprisingly, we find that the initial mass of planetesimals has little influence on the output in our models, which we will discuss in \se{ini_mass}.

\subsection{Pebble accretion}\label{sec:PA}
\hjrem{Pebble accretion as a mechanism driving planet formation has seen active development in the last decade 
. However, i} In a smooth disc without substructures, in order to trigger efficient pebble accretion at the outer region of the disc ($>50$ au), a ${\sim}1000$~km size embryo is required \citep{VisserOrmel2016}. Thanks to the high concentration of pebbles inside the ring, the threshold of \hjch{PA}{pebble accretion} is significantly lower.

We calculate the pebble accretion rate of each planetesimal \hjch{depending n the surface density of the pebble and}{at} the semimajor axis  \hjch{of}{$a_{\rm p}$}
\begin{equation}\label{eq:dotMPA}
    \dot{M}_{\rm PA} = {\cal R} \Sigma_{\rm peb} a_{\rm p}^2 \Omega_K
\end{equation}
where $\Sigma_{\rm peb}$ is the local pebble surface density. The accretion efficiencies in the 2D and 3D limits read \citep[converted from][]{LiuOrmel2018,OrmelLiu2018}
\begin{subequations}\label{eq:RPA}
\begin{align}
    {\cal R} &= ({\cal R}_{\rm 2D}^{-2} + {\cal R}_{\rm 3D}^{-2})^{-0.5}    \\
    {\cal R}_{\rm 2D} &= 4.0 f_{\rm set} \sqrt{\frac{q{\rm St}\Delta v}{v_K}}
    \label{Za}    \\
    {\cal R}_{\rm 3D} &= 4.9 f_{\rm set}^2 \frac{q{\rm St}}{h_{\rm peb}}
    \label{Zb}
\end{align}
\end{subequations}
in which $q = m/M_\star$ is the mass ratio of planetesimal and $\Delta v$ (\eq{delv}) is the approaching velocity of pebble relative to the planetesimal. The settling factor $f_{\rm set}$ is always around unity. Detailed derivation and formulas are \hjadd{listed} in \App{PAE}. Since planetesimals with high inclination $i > h_{\rm d}$ only interact with pebbles over a fraction $~i/h_{\rm d}$ of their orbits. The pebble scaleheight coefficient should be modified as
\begin{equation}\label{eq:hpeb}
    h_{\rm peb} = \sqrt{h_{\rm d}^2 + \frac{\pi i^2}{2}\left(1-\exp{\left[-\frac{i}{2h_{\rm d}}\right]}\right)}
\end{equation}
where $i$ is the inclination and 
\begin{equation}
    h_{\rm d} \equiv \frac{H_{\rm d}}{r} =
    \left\{
    \begin{array}{lr}
        \frac{H_{\rm p}}{r} & \rm PB \\
        \frac{H_{\rm c}}{r} & \rm CR
    \end{array}
    \right.
\end{equation}
For the PB model, the planetesimal typically has $i \ll h_{\rm d}$\hjch{.}{,} while for the CR model the \hjch{punishment}{reduction of the accretion rate} due to the inclination \hjch{on the pebble accretion rate}{of the planetesimals} could be\hjadd{come} significant because of the thin pebble layer.

In our simulations, we calculate the pebble accretion rate for each planetesimal. Thus the total mass flux of $N$ planetesimals accreting from the ring is
\begin{equation}
\label{eq:dotMPA_ring}
    \dot{M}_{\rm PA,ring} = -\sum_{i}^{N} \dot{M}_{\rm PA,i}
\end{equation}

For both PB and CR, we solve for the ring mass semi-analytically. We set an external mass flux $\dot{M}_{\rm ext}$, which is the total mass flux of pebbles drifting from the outer disc to feed the ring location. 
\hjadd{}
For PB runs, we assume that the amplitude of the pressure bump is strong enough to halt all of the pebbles---a perfect dust filtration \citep[e.g.,][]{ZhuEtal2012}. However, since there is no pressure support to stop the radial drift, in the CR discs, the rings always leak pebbles downstream. Following JO21, the leaking mass flux is
\begin{equation}\label{eq:dotM_leak}
\begin{aligned}
    \dot{M}_\mathrm{leak}
    &\simeq \mathrm{St} \Sigma_\mathrm{g,0} h_0^3 M_\star^{0.5} r^{0.25} \\ 
    & = 50
    \left(\frac{\mathrm{St}}{10^{-2}}\right)
    \left(\frac{\Sigma_\mathrm{g,0}}{3.6\,\mathrm{g}\,\mathrm{cm}^{-2}}\right)
    \left(\frac{h_0}{0.07}\right)^3 \\
    & \times
    \left(\frac{M_\star}{M_\odot}\right)^{0.5} \left(\frac{r}{r_0}\right)^{0.25}
    \,M_\oplus\,\mathrm{Myr}^{-1}
\end{aligned}
\end{equation}
This means that a moderate amount of pebbles can leak from the ring toward the host star, which might further promote the growth of the embryos or the ring in the inner disc. For completeness, we account for the pebble accretion from the smooth pebble background disc away from the ring. We introduce a background factor
\begin{equation}\label{eq:f_bkg}
    f_{\rm bkg} = \frac{1}{2}(1+\tanh[\frac{2(|r_0-r|-2w_{\rm ring})}{w_{\rm ring}}])
\end{equation}
which is zero within one $w_{\rm ring}$ width of the ring centre, and will increase to unity rapidly when $>3 w_{\rm ring}$ away from the central location of the ring. \Fg{f_bkg} show the radial distribution of the $f_{\rm bkg}$ factor. The pebble flux upstream is $\dot{M}_{\rm ext}$ and the pebble flux downstream is $\dot{M}_{\rm leak}$. For PB rings, planetesimal may only be able to feed from the upstream flux. With the the given pebble flux, we simply use the pebble accretion efficiency $\epsilon_\mathrm{PA}$ to calculate the accretion rate (see \eq{eps2d} and \eq{eps3d}). The result is then
\begin{equation}
\label{eq:mPA_disc}
    \dot{M}_{\rm PA, disc} \equiv
    \left\{
    \begin{array}{lr}
        \dot{M}_{\rm PA, up} = -f_{\rm bkg}\epsilon_\mathrm{PA} \dot{M}_{\rm ext} & \rm r>r_0   \\
        \dot{M}_{\rm PA, down} = -f_{\rm bkg}\epsilon_\mathrm{PA} \dot{M}_{\rm leak} & r<r_0,\,\rm CR  \\
        \dot{M}_{\rm PA, down} = 0 & r<r_0,\,\rm PB
    \end{array}
    \right.
\end{equation}
In most of the simulations, $\dot{M}_{\rm PA, disc}$ is generally very low compared with $\dot{M}_{\rm PA, ring}$ in \eq{dotMPA_ring}. Because of mass conservation, the mass flux that eventually feeds the ring is then
\begin{equation}
    \dot{M}_{\rm ring} = \dot{M}_{\rm ext} - \dot{M}_{\rm leak} - \sum_{i}^{N} \dot{M}_{\rm PA,up,i}.
\end{equation}
\hjch{Meanwhile}{In addition}, the ring \hjrem{also}loses mass by planetesimal formation \eq{dotM_plt} and pebble accretion \eq{dotMPA_ring}. We integrate the ring mass by calculating the mass flux change in the N-body simulation. We then change the ring mass \eq{M_ring} and the pebble surface density at the ring peak \eq{Sigr} accordingly.

\begin{figure}
    \centering
    \includegraphics[width=0.99\columnwidth]{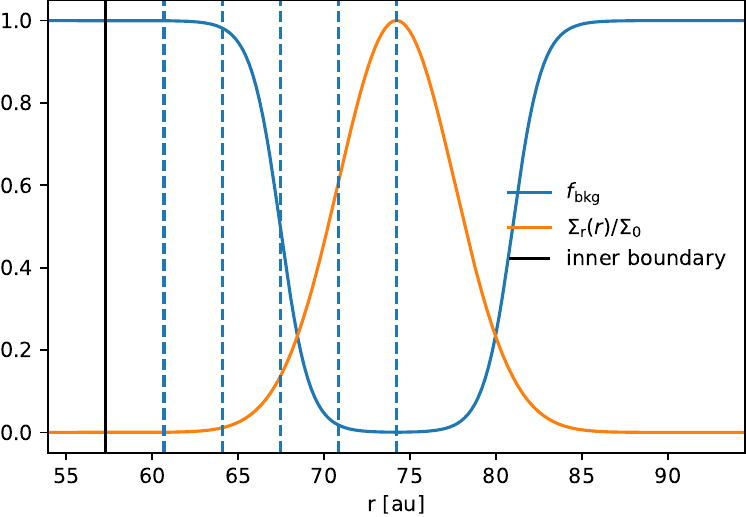}
    \caption{\label{fig:f_bkg} Radial distribution of the $f_{\rm bkg}$ factor (orange line). Orange line indicates the radial profile of the ring's surface density. The separation of two neighbored blue dash lines is $w_{\rm ring}$.}
\end{figure}

\subsection{Treatment of planet migration}\label{sec:pl_mg}
Massive bodies feel the gravitational torques from the disc gas \citep[so called type I migration, e.g.][]{GoldreichTremaine1979,KleyNelson2012,Baruteauetal2014}. We implement the following expressions in the N-body code as accelerations experienced by the planets due to type I migration torque \citep[e.g.,][]{PapaloizouLarwood2000}
\begin{equation}
\begin{aligned}
    \bm{a}_{\rm m} &= -\frac{\bm{v}}{t_{\rm mg}} \\
    \bm{a}_{\rm e} &= -2\frac{(\bm{v}\cdot\bm{r})\bm{r}}{r^2t_{\rm e}} \\
    \bm{a}_{\rm i} &= -\frac{\bm{v}_{\rm z}}{t_{\rm i}}
\end{aligned}
\end{equation}
where $\bm{v}=(v_x, v_y, v_z)$ and $\bm{r}=(x, y, z)$ are the velocity and position of the planet. The characteristic migration timescale $t_{\rm mg}$ for a planet on a circular orbit is \citep[e.g.,][]{CresswellNelson2008} 
\begin{equation}\label{eq:t_mg}
\begin{aligned}
    t_{\rm mg} &= \frac{1}{2}
    \frac{1}{f_{\rm mg}}\frac{1}{q}\frac{M_\star }{\Sigma_{\rm g} a_{\rm p}^2} h_{\rm g}^2 \frac{1}{\Omega_K}    \\
    &= \frac{40\,\mathrm{Myr}}{f_\mathrm{mg}} 
    \left(\frac{m_{\rm p}}{1\,M_\oplus}\right)^{-1} \left(\frac{\Sigma_{\rm g,0}}{3.6\,\rm g\,cm^{-2}}\right)^{-1} \left(\frac{h_0}{0.07}\right)^2 \left(\frac{a_{\rm p}}{r_0}\right)
\end{aligned}
\end{equation}
where $a_{\rm p}$ is the semimajor axis of the planet and $f_{\rm mg}$ the type I migration prefactor \hjadd{(see below).  We neglect the aerodynamic gas drag force on the planetesimals and planet, since the damping caused by it is inefficient for large planetesimals, and strongly decreases as a function of the semimajor axis \citep[][]{AdachiEtal1976}. Yet, the tidal interactions (type I torque) between a planetary embryo and gas also damps its eccentricity and inclination. Similar to \eq{t_mg},}
\begin{equation}\label{eq:t_ei}
\begin{aligned}
    t_{\rm e} &= t_{\rm i} = \frac{1}{q}\frac{M_\star }{\Sigma_{\rm g} a_{\rm p}^2} h_{\rm g}^4 \frac{1}{\Omega_K}    \\
    &= 0.4 \times \left(\frac{m_{\rm p}}{1\,M_\oplus}\right)^{-1} \left(\frac{\Sigma_{\rm g,0}}{3.6\,\rm g\,cm^{-2}}\right)^{-1} \left(\frac{h_0}{0.07}\right)^4 \left(\frac{a_{\rm p}}{r_0}\right)\,\rm Myr
\end{aligned}
\end{equation}
are the eccentricity, and inclination damping timescales.

\hjadd{For migration, we consider three setups:
\begin{enumerate}
    \item Simple, inwards migration. We take $f_{\rm mg}=1$ in most of our simulations but also test different values $(f_\mathrm{mg}>0)$ for the parameter variation in the CR model. In this setup, a pressure bump only affects pebbles (\eq{pr}), but \textit{not} the planet migration (\eq{t_mg}). This choice is inconsistent -- a change in pressure after all must be reflected in either density, temperature or both -- but it serves the purpose of comparing the CR and PB models.
    \item No migration, $f_\mathrm{mg}\to0$. 
\end{enumerate}
And for the pressure bump model:
\begin{enumerate}
    \setcounter{enumi}{2}
    \item Realistic migration (\se{real_bump}). We take $f_\mathrm{mg} = (2.7+1.1\beta)/3.9$ consistent with the locally isothermal disc model  \citep{CresswellNelson2008}. In \se{real_bump} we explicitly specify the origin of the pressure bump to be either a maximum in temperature \citep[e.g.,][]{KimEtal2020,RosottiEtal2021} or density  \citep[e.g.,][]{MassetEtal2006,Morbidelli2020}. Consequently, this affects the values of $\Sigma_{\rm g}$, $h_{\rm g}$, $\eta$ and the surface density gradient $\beta$ (as in $\Sigma_{\rm g}(r) \propto r^{-\beta}$). In the case of a density origin, the direction of migration can be reversed at so-called trapping locations.  
\end{enumerate}}

\hjadd{For simplicity, we neglect other contributions to the total torque like the heating torques induced by temperature rise associated with infalling pebble \citep[e.g.,][]{Benitez-LlambayEtal2015,Masset2017,GuileraEtal2019,GuileraEtal2021} and torques induced by scattered pebble flow \citep[][]{Benitez-LlambayPessah2018}, both of which can hinder the inward migration of low-mass planet and lead the results closer to a migration trapping cases. And we also ignore the influence caused by non-isothermal effects, e.g., thermal and mass diffusivity, cooling, and temperature gradient \citep[e.g.,][]{PaardekooperEtal2010,PaardekooperEtal2011}.}

\subsection{Other numerical setups}
\subsubsection{Initial setup of the ring}
We initialize the ring mass such that the mid-plane dust-to-gas ratio $Z = 1$ at the ring peak 
\begin{equation}
    \Sigma_{\rm r,0} = \Sigma_{\rm g,0} \frac{H_{\rm d}}{H_{\rm g}}
\end{equation}
The motivation of this setup is simply because that there is no planetesimal formation when the dust-to-gas ratio is below unity, and all of the mass flux can be fed to the ring. We estimate the timescale required for the ring to grow to the initial mass by $t_{\rm load} = M_{\rm ring,0}/\dot{M}_{\rm ring}$, whose value is listed in \tb{output} for each run.

\subsubsection{Planetesimal removal}

The excitation of the eccentricity and inclination of planetesimals formed in the ring is mainly due to planetesimals' viscous stirring by themselves. As we derived, highly excited planetesimal will have a low pebble accretion rate because of their high relative velocity towards pebble and the low pebble density out of the ring region. These small planetesimals contribute only a small fraction of the total mass, and grow little in long time evolution but dominate in the number of planetesimals, which significantly slows down the N-body simulations. 

In order to improve the computational efficiency, we remove planetesimals that are scattered away from the ring and therefore cannot accrete efficiently. We introduce a space factor $f_{\rm space}$, which is the area that the particle's orbit overlap with the ring region in $r-z$ plane
\begin{equation}
\begin{aligned}
    f_{\rm space} &=\dfrac{\min((1+e)a_{\rm p}, r_0+w_{\rm ring})-\max((1-e)a_{\rm p}, r_0-w_{\rm ring})}{2 e a_{\rm p}}    \\
    &\times \min(1, \frac{h_{\rm d}}{i a_{\rm p}})
\end{aligned}
\end{equation}
For planetesimals whose orbits fall in the cube centred at the centre the ring with length $2w_{\rm ring}$ and height $h_{\rm d}$, the space factor $f_{\rm space} = 1$. We remove the particle only when its $f_{\rm space}$ falls below 0.1 \textit{and} the mass is lower than Mars. Planetesimals that are currently inside the ring region will not be removed. In addition, we rank the planetesimals by mass and never remove those whose mass ranks in the top $N_{\rm min} = 100$. We also test the case $N_{\rm min} = 200, 300, 500, 1000$ for our default model and we find that our results are insensitive to the choice of $N_{\rm min}$.

\hspace*{\fill}

We set an inner boundary $r_{\rm in} = r_0-5w_{\rm ring}$ for our simulation with type-I migration on, i.e., planets who migrate and pass through the inner boundary will be removed. We discuss the post-ring evolution of these planets in \se{post_ring}.

The simulations were conducted using the \texttt{IAS15} integrator of the open-source N-body code \texttt{REBOUND}, with collisions between bodies treated as perfect mergers that conserve the linear momentum \citep{ReinLiu2012}. The pebble accretion rate expressions and type-I migration force were implemented via \texttt{REBOUNDx} \citep{TamayoEtal2020}.

\section{Results}\label{sec:results}
The aim of this work is to test the viability of planet formation in planetesimal-forming rings. Without loss of generality, we pick the default parameters similar to the values used in JO21 and choose the ring location and shape the same as the B77 ring in AS~209. We summarize these default parameters in \tb{pars} in bold. Different setups between the CR and PB runs are also listed. In \tb{input}, we \hjch{give}{list} all runs and the corresponding key parameters studied in each.

\begin{table}
\caption{Summary of parameters studied and comparison between the Clumpy Ring (CR) and the Pressure-Bump-induced ring (PB).
\label{tab:pars}}
\centering
\small
\begin{tabular}{l|cc}
\hline\hline
name & Clumpy Ring (CR) & Pressure Bump (PB) \\
\hline
\multicolumn{3}{l}{Same}    \\
\hline
$r_0$ & \multicolumn{2}{c}{30, {\bf 74.2}, 154~au} \\
$w_{\rm ring}$ & \multicolumn{2}{c}{\textbf{3.38~au}} \\
$m_{\rm 0}$ & \multicolumn{2}{c}{$M_{\rm Ceres}$, \textbf{$M_{\rm Pluto}$}, $M_{\rm Moon}$, $M_{\rm Mars}$} \\
$\dot{M}_{\rm ext}$ & \multicolumn{2}{c}{60, 80, {\bf 100}, 150, 300 $M_\oplus\,\rm Myr^{-1}$} \\
$\delta_{\rm z}$ & \multicolumn{2}{c}{1e-4, {\bf 1e-3}} \\
St & \multicolumn{2}{c}{1e-3, 3e-3, {\bf 1e-2}, 3e-3, 1e-1} \\
$\zeta$ & \multicolumn{2}{c}{1e-5, 1e-4, {\bf 1e-3}, 1e-2} \\
\hline
\multicolumn{3}{l}{Difference}  \\
\hline
$w_{\rm pb}$ & - & $\sqrt{\frac{\rm St}{\alpha}} w_d$\\
$H_{\rm d}$ & $\sim\eta r$ & $\sqrt{\frac{\delta_{\rm z}}{\rm St}}H_{\rm g}$ \\
$\dot{M}_{\rm leak}$ & \eq{dotM_leak} & 0 \\
$\eta$ & \eq{eta} & \eq{eta_pb} \\
\hline
\end{tabular}
\end{table}

\begin{table}
\caption{Model runs input parameters.\label{tab:input} (1) Stokes number; (2) external mass flux; (3) leaking mass flux \eq{dotM_leak}; (4) note on the focus of parameter studies.}
\centering
\small
\begin{tabular}{l|ccc|l}
\hline\hline
run-id$^a$ & St & $\dot{M}_{\rm ext}$ & $\dot{M}_{\rm leak}$ & note \\
& & \multicolumn2c{[$M_\oplus\rm\,Myr^{-1}$]} & \\
& (1) & (2) & (3) & (4)\\
\hline
\multicolumn{4}{l}{runs varying initial planetesimal mass} & $m_{\rm 0}$   \\
\texttt{cr-mMars} & 0.01 & 100 & 50 & $ M_{\rm Mars}$\\
\texttt{cr-mMoon} & 0.01 & 100 & 50 & $M_{\rm Moon}$\\
\texttt{cr-default} & 0.01 & 100 & 50 & $M_{\rm Pluto}$\\
\texttt{cr-m05Pluto} & 0.01 & 100 & 50 & $0.5\,M_{\rm Pluto}$\\
\texttt{cr-m03Pluto} & 0.01 & 100 & 50 & $0.3\,M_{\rm Pluto}$\\
\texttt{cr-m02Pluto} & 0.01 & 100 & 50 & $0.2\,M_{\rm Pluto}$\\
\texttt{cr-mCeres} & 0.01 & 100 & 50 & $M_{\rm Ceres}$\\
\texttt{pb-mMars} & 0.01 & 100 & 0 & $ M_{\rm Mars}$\\
\texttt{pb-mMoon} & 0.01 & 100 & 0 & $M_{\rm Moon}$\\
\texttt{pb-default} & 0.01 & 100 & 0 & $M_{\rm Pluto}$\\

\hline
\texttt{cr-F60}  & 0.01 & 60 & 50 & \\
\texttt{cr-F80}  & 0.01 & 80 & 50 & \\
\texttt{cr-F150} & 0.01 & 150 & 50 & \\
\texttt{cr-F300} & 0.01 & 300 & 50 & \\

\hline
\texttt{cr-St001} & 0.001 & 55  & 5 & \\
\texttt{cr-St003} & 0.003 & 65  & 15 & \\
\texttt{cr-St030} & 0.03  & 200 & 150 & \\
\texttt{cr-St100} & 0.1   & 550 & 500 & \\
\texttt{pb-St030} & 0.03  & 100 & 0 & \\
\texttt{pb-St100} & 0.1   & 100 & 0 & \\

\hline
\texttt{cr-St001F100} & 0.001 & 100 & 5 & \\
\texttt{cr-St003F100} & 0.003 & 100 & 15 & \\
\texttt{cr-St006F100} & 0.006 & 100 & 30 & \\
\texttt{cr-St015F100} & 0.015 & 100 & 75 & \\
\texttt{cr-St018F100} & 0.018 & 100 & 90 & \\

\hline
\multicolumn{4}{l}{runs varying planetesimal formation efficiency} & $\zeta$ \\
\texttt{cr-e2} & 0.01 & 100 & 50 & $10^{-2}$\\
\texttt{cr-default} & 0.01 & 100 & 50 & $10^{-3}$\\
\texttt{cr-e4} & 0.01 & 100 & 50 & $10^{-4}$\\
\texttt{cr-e5} & 0.01 & 100 & 50 & $10^{-5}$\\

\hline
\multicolumn{4}{l}{runs varying vertical dust diffusivity} & $\delta_{\rm z}$ \\
\texttt{cr-default} & 0.01 & 100 & 50 & $10^{-3}$\\
\texttt{cr-a4} & 0.01 & 100 & 50 & $10^{-4}$\\
\texttt{pb-default} & 0.01 & 100 &  0 & $10^{-3}$\\
\texttt{pb-a4} & 0.01 & 100 &  0 & $10^{-4}$\\


\hline
\multicolumn{4}{l}{runs varying vertical ring location} & $r_0$ \\
\texttt{cr-r30}  & 0.01 & 100 & 40 & 30~au\\
\texttt{cr-default} & 0.01 & 100 & 50 & 74.2~au\\
\texttt{cr-r154} & 0.01 & 100 & 60 & 154~au\\

\hline
\texttt{cr-05mg} & 0.01 & 100 & 50 & $f_{\rm mg}=0.5$\\
\texttt{cr-20mg} & 0.01 & 100 & 50 & $f_{\rm mg}=2$\\
\texttt{cr-nmg} & 0.01 & 100 & 50 & $f_{\rm mg}\to0$\\
\texttt{pb-nmg} & 0.01 & 100 & 0  & $f_{\rm mg}\to0$\\
\texttt{pb-St030-nmg} & 0.03 & 100 & 0  & $f_{\rm mg}\to0$\\
\texttt{pb-St100-nmg} & 0.1  & 100 & 0  & $f_{\rm mg}\to0$\\

\hline
\texttt{pb-t-bump} & 0.01 & 100 &  0 & temperature bump\\
\texttt{pb-d-bump} & 0.01 & 100 &  0 & gas density bump\\

\hline
\\

\end{tabular}
\end{table}


\begin{table*}
\caption{Model runs results.\label{tab:output} (1) time required to load the ring mass to $\rho_{\rm d}/\rho_{\rm g}=1$ by the external mass flux; (2) time when the first protoplanet leaves the inner simulation domain; (3) average time interval between protoplanets leaving the box; error bars indicate the maximum and minimum intervals duration; (4) time when the sixth protoplanet passes the inner simulation domain; (5) the mass of the biggest protoplanet; (6) the average mass of protoplanet passing the inner simulation domain; (7) the predicted representative planet mass \eq{m_rpr}; (8) the initial mass of the ring; (9) the measured averaged ring mass in the saturation stage; (10) the predicted saturated mass of the ring \eq{M_ring_ss}; (11) section reference where the model is mainly discussed.}
\centering
\small
\begin{tabular}{l|ccccc|ccccc|l}
\hline\hline
run-id$^a$ & $t_{\rm load}$ & $t_{\rm 1}$ & $\langle\Delta\,t \rangle$ & $t_{\rm 6}$& $m_{\rm max}$ & $m_{\rm avg}$ & $m_{\rm rp}$ & $m_{\rm ring,0}$ & $m_{\rm ring,ms}$ & $m_{\rm ring,ss}$ & Section\\
&  \multicolumn4c{[Myr]} & \multicolumn6c{[$M_\oplus$]} \\
& (1) & (2) & (3) & (4) & (5) & (6)& (7) & (8) & (9) & (10) & (11)\\
\hline
\texttt{cr-mCeres}   & 0.23 & 1.06 & $0.27^{+0.28}_{-0.26}$ & 2.44 & 15.1 & 11.8 & 18 & 11.6 & 34.8 & 34 & \sef{ini_mass} \\
\texttt{cr-m02Pluto} & 0.23 & 1.04 & $0.34^{+0.43}_{-0.27}$ & 2.73 & 21.7 & 12.1 & 18 & 11.6 & 33.1 & 34 & \sef{ini_mass} \\
\texttt{cr-m03Pluto} & 0.23 & 0.97 & $0.17^{+0.31}_{-0.16}$ & 1.83 & 21.5 & 10.6 & 18 & 11.6 & 31.6 & 34 & \sef{ini_mass} \\
\texttt{cr-m05Pluto} & 0.23 & 1.01 & $0.14^{+0.25}_{-0.14}$ & 1.73 & 13.5 &  8.3 & 18 & 11.6 & 30.7 & 34 & \sef{ini_mass} \\
\texttt{cr-default}  & 0.23 & 0.97 & $0.18^{+0.33}_{-0.17}$ & 1.87 & 22.2 & 10.6 & 18 & 11.6 & 30.5 & 34 & \sef{default_model}\\
\texttt{cr-mMoon}    & 0.23 & 0.85 & $0.35^{+0.29}_{-0.26}$ & 2.61 & 19.2 & 13.4 & 18 & 11.6 & 32.6 & 34 & \sef{ini_mass} \\
\texttt{cr-mMars}    & 0.23 & 0.79 & $0.28^{+0.33}_{-0.26}$ & 2.18 & 17.9 & 10.7 & 18 & 11.6 & 28.2 & 34 & \sef{ini_mass} \\

\texttt{pb-default}  & 1.61 & 0.65 & $0.04^{+0.07}_{-0.04}$ & 0.87 & 19.7 &  6.9 & 25 & 160.5 & 141.1 & 54 & \sef{SDCRPB} \\
\texttt{pb-mMoon}    & 1.61 & 0.61 & $0.04^{+0.06}_{-0.03}$ & 0.80 &  9.8 &  5.3 & 25 & 160.5 & 152.4 & 54 & \sef{ini_mass} \\
\texttt{pb-mMars}    & 1.61 & 0.49 & $0.20^{+0.18}_{-0.15}$ & 1.51 & 33.6 & 17.3 & 25 & 160.5 & 167.4 & 54 & \sef{ini_mass} \\

\hline
\texttt{cr-F60}     & 1.16 & 1.71 & $0.56^{+0.84}_{-0.53}$ & 4.50 &  9.0 &  5.6 &  8 & 11.6 & 13.9 & 12 & \sef{dotM_St} \\
\texttt{cr-F80}     & 0.39 & 1.24 & $0.47^{+0.59}_{-0.36}$ & 3.60 & 21.5 & 11.3 & 14 & 11.6 & 23.9 & 24 & \sef{dotM_St} \\
\texttt{cr-default} & 0.23 & 0.97 & $0.18^{+0.33}_{-0.17}$ & 1.87 & 22.2 & 10.6 & 18 & 11.6 & 30.5 & 34 & \sef{dotM_St} \\
\texttt{cr-F150}    & 0.12 & 0.77 & $0.22^{+0.26}_{-0.21}$ & 1.85 & 26.1 & 17.5 & 26 & 11.6 & 46.2 & 53 & \sef{dotM_St} \\
\texttt{cr-F300}    & 0.05 & 0.58 & $0.04^{+0.10}_{-0.04}$ & 0.79 & 29.1 &  9.7 & 41 & 11.6 & 92.4 & 99 & \sef{dotM_St} \\

\hline
\texttt{cr-St001}   & 0.23 & 2.31 & $0.56^{+0.69}_{-0.43}$ & 5.12 & 11.9 &  7.9 & 18 & 11.6 &126.1& 156& \sef{dotM_St} \\
\texttt{cr-St003}   & 0.23 & 1.40 & $0.38^{+0.28}_{-0.30}$ & 3.32 & 19.7 &  8.0 & 18 & 11.6 & 82.5 & 75 & \sef{dotM_St} \\
\texttt{cr-default} & 0.23 & 0.97 & $0.18^{+0.33}_{-0.17}$ & 1.87 & 22.2 & 10.6 & 18 & 11.6 & 30.5 & 34 & \sef{dotM_St} \\
\texttt{cr-St030}   & 0.23 & 0.89 & $0.22^{+0.40}_{-0.19}$ & 1.99 & 28.1 & 13.2 & 18 & 11.6 & 12.6 & 16 & \sef{dotM_St} \\
\texttt{cr-St100}   & 0.23 & 0.69 & $0.12^{+0.12}_{-0.10}$ & 1.29 & 23.4 & 13.1 & 18 & 11.6 &  5.0 &  7 & \sef{dotM_St} \\

\texttt{pb-St030}   & 0.96 & 0.29 & $0.06^{+0.05}_{-0.04}$ & 0.57 & 14.9 &  8.6 & 25 & 95.6 & 39.0 & 26 & \sef{SDCRPB} \\
\texttt{pb-St100}   & 0.53 & 0.23 & $0.13^{+0.30}_{-0.12}$ & 0.87 & 30.1 & 14.0 & 25 & 53.0 & 16.4 & 12 & \sef{SDCRPB} \\

\hline
\texttt{cr-St001F100} & 0.12 & 1.71 & $0.43^{+0.37}_{-0.26}$ & 3.88 & 24.2 & 10.4 & 25 & 11.6 & 216.3 & 240 & \sef{dotM_St} \\
\texttt{cr-St003F100} & 0.14 & 1.28 & $0.25^{+0.28}_{-0.14}$ & 2.52 & 27.1 & 12.8 & 24 & 11.6 & 109.3 & 107 & \sef{dotM_St} \\
\texttt{cr-St006F100} & 0.17 & 1.00 & $0.33^{+0.45}_{-0.26}$ & 2.65 & 32.3 & 15.0 & 22 & 11.6 &  60.2 &  59 & \sef{dotM_St} \\
\texttt{cr-default}   & 0.23 & 0.97 & $0.18^{+0.33}_{-0.17}$ & 1.87 & 22.2 & 10.6 & 18 & 11.6 &  30.5 &  34 & \sef{dotM_St} \\
\texttt{cr-St015F100} & 0.46 & 1.02 & $0.40^{+0.24}_{-0.29}$ & 3.01 & 13.2 & 11.0 & 13 & 11.6 &  14.5 &  16 & \sef{dotM_St} \\
\texttt{cr-St018F100} & 2.32 & 1.15 & $0.51^{+0.95}_{-0.51}$ & 3.69 & 10.2 &  7.0 &  8 & 11.6 &     8.8 &   8  & \sef{dotM_St} \\

\hline
\texttt{cr-e2}      & 0.23 & 0.93 & $0.39^{+0.27}_{-0.39}$ & 2.90 & 13.2 &  9.6 & 18 & 11.6 & 22.1 & 34 & \sef{epsilon} \\
\texttt{cr-default} & 0.23 & 0.97 & $0.18^{+0.33}_{-0.17}$ & 1.87 & 22.2 & 10.6 & 18 & 11.6 & 30.5 & 34 & \sef{epsilon} \\
\texttt{cr-e4}      & 0.23 & 0.91 & $0.17^{+0.31}_{-0.16}$ & 1.74 & 19.6 &  9.5 & 18 & 11.6 & 32.0 & 34 & \sef{epsilon} \\
\texttt{cr-e5}      & 0.23 & 0.97 & $0.16^{+0.35}_{-0.16}$ & 1.78 & 22.0 & 12.0 & 18 & 11.6 & 35.6 & 34 & \sef{epsilon} \\

\hline
\texttt{cr-default} & 0.23 & 0.97 & $0.18^{+0.33}_{-0.17}$ & 1.87 & 22.2 & 10.6 & 18 &  11.6 & 30.5 & 34 & \sef{default_model}\\
\texttt{cr-a4}      & 0.23 & 0.74 & $0.09^{+0.07}_{-0.09}$ & 1.19 & 50.7 & 23.5 & 18 &  11.6 & 23.7 & 34 & \sef{post_ring} \\
\texttt{pb-default} & 1.61 & 0.65 & $0.04^{+0.07}_{-0.04}$ & 0.87 & 19.7 &  6.9 & 25 & 160.5 & 141.1 & 54 & \sef{SDCRPB} \\
\texttt{pb-a4}      & 1.61 & 0.63 & $0.14^{+0.33}_{-0.13}$ & 1.32 & 59.3 & 28.5 & 25 &  53.0 &  59.0 & 54 & \sef{SDCRPB} \\


\hline
\texttt{cr-r30}     & 0.17 & 0.90 & $0.16^{+0.34}_{-0.14}$ & 1.71 & 17.2 & 10.0 & 20 & 10.4 & 15.5 & 24 & \sef{r0} \\
\texttt{cr-default} & 0.23 & 0.97 & $0.18^{+0.33}_{-0.17}$ & 1.87 & 22.2 & 10.6 & 18 & 11.6 & 30.5 & 34 & \sef{default_model} \\
\texttt{cr-r154}    & 0.30 & 1.23 & $0.26^{+0.34}_{-0.16}$ & 2.51 & 22.8 &  8.3 & 16 & 12.3 & 43.2 & 42 & \sef{r0} \\

\hline
\texttt{cr-05mg}      & 0.23 & 1.20 & $0.40^{+0.30}_{-0.27}$ & 3.21 & 26.9 & 14.0 & 25 & 11.6 & 24.7 & 27 & \sef{nmg} \\
\texttt{cr-20mg}      & 0.23 & 0.78 & $0.22^{+0.16}_{-0.16}$ & 1.89 & 13.0 &  8.1 & 13 & 11.6 & 40.5 & 43 & \sef{nmg} \\
\hline
\texttt{cr-nmg}       & 0.23 & 1.57$^a$ & & & 60$^b$ & & & 11.6 &  11.0$^c$ & & \sef{nmg} \\
\texttt{pb-nmg}       & 1.61 & 1.60 & & & 60 & & &160.5 & 155.1 & & \sef{nmg} \\
\texttt{pb-St030-nmg} & 0.96 & 0.74 & & & 60 & & & 95.6 &  51.3 & & \sef{nmg} \\
\texttt{pb-St100-nmg} & 0.53 & 0.84 & & & 60 & & & 53.0 &  16.6 & & \sef{nmg} \\

\hline
\texttt{pb-t-bump} & 1.61 & 0.61 & $0.07^{+0.04}_{-0.05}$ & 0.96 & 29.6 &  10.4 & 25 & 160.5 & 134.2 & 54 & \sef{real_bump} \\
\texttt{pb-d-bump}    & 1.61 & 0.79 & & & 60 & & &160.5 & 119.1 & & \sef{real_bump} \\

\hline
\multicolumn{12}{l}{\footnotesize$^a$ since there is no migration, $t_{\rm 1}$ is defined as the time when the most massive planet open a gap, see \se{nmg}}\\
\multicolumn{3}{l}{\footnotesize$^b$ gap-opening mass \eq{M_gap}}
& \multicolumn{9}{l}{\footnotesize$^c$ the mass of the ring at $t_{\rm 1}$, the ring should eaten by the planet eventually}\\

\end{tabular}
\end{table*}

\subsection{Default model}\label{sec:default_model}
\hjadd{In }\fg{a_t_default}\hjadd{, we} shows the trajectories of protoplanets that grow to a mass higher than $1.7\,M_\oplus$ during the simulation \hjadd{in run \texttt{cr-default}}. \hjadd{The colour of the trajectory line indicates the inclination of the embryos. The cyan background corresponds to the density of planetesimals, in which the protoplanet (trajectory lines) are not included.} We run the simulation until $3\,\mathrm{Myr}$, which is long enough to spawn ten protoplanets as shown in the \fg{a_t_default}. The criterion of $1.7\,M_\oplus$ is chosen so that the planets are massive enough to migrate inward sufficiently. According to \eq{t_mg}, a planet as massive as $1.7\,M_\oplus$ can migrate over a distance equal to the ring width in $w_{\rm ring}/r_0 \times t_{\rm mg} = 1\,\mathrm{Myr}$. Finally, nine planets have passed the inner boundary of the simulation box. We pick the first six as the sample for further statistic study. In order to compare the default model with other runs equally, we continue each simulation until six planets have crossed the inner domain boundary. This number is sufficient to obtain statistically viable results while limiting the computational cost.

\begin{figure}
    \centering
    \includegraphics[width=0.99\columnwidth]{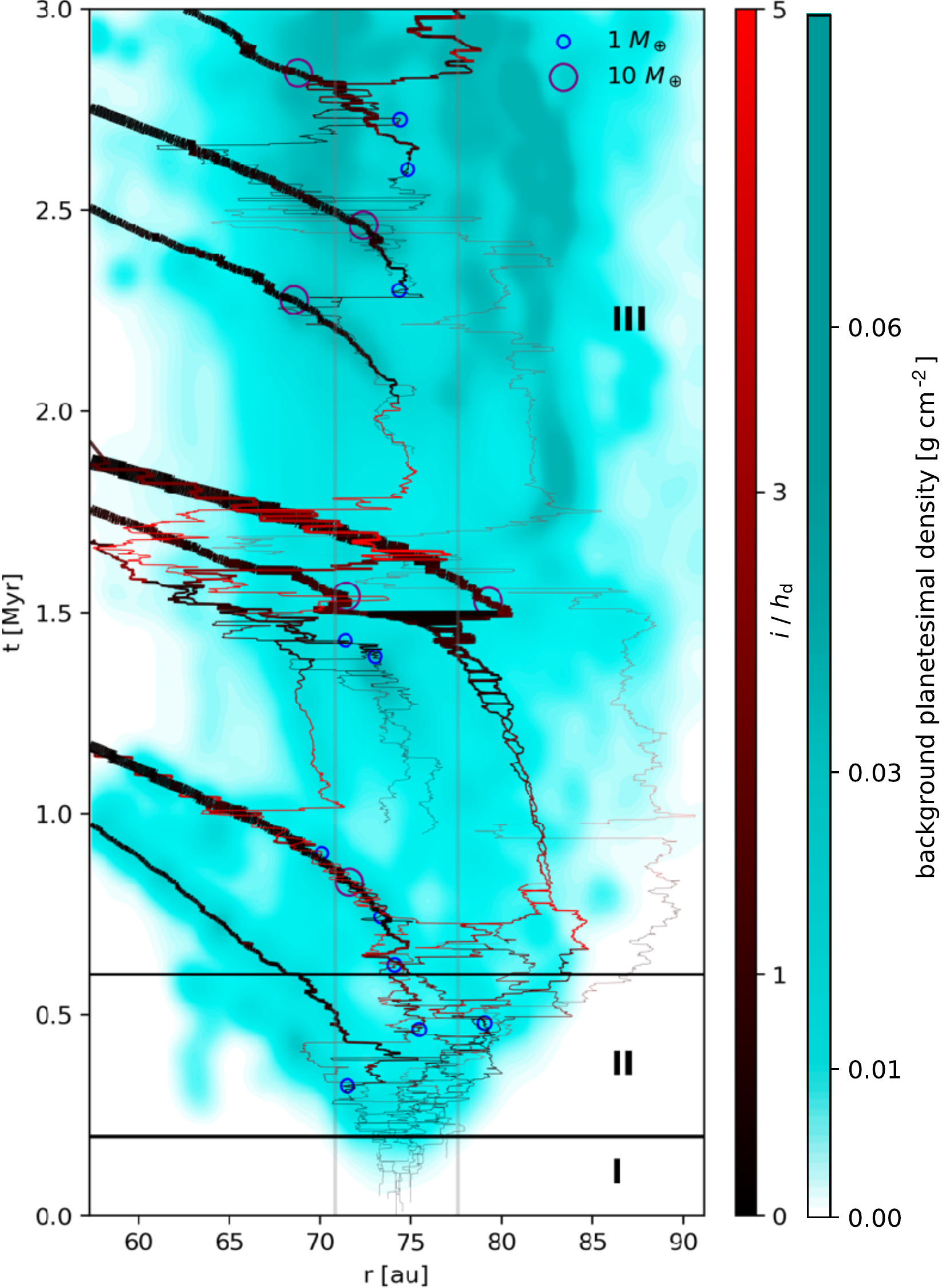}
    \caption{\label{fig:a_t_default} Protoplanet trajectories (solid lines) of the default model \texttt{cr-default}. The colour of the solid lines corresponds to the inclination of the protoplanets. The line width is linear with the mass of the planet. \hjadd{We mark the planets with small (big) circles when they reach $1\,M_\oplus$ ($10\,M_\oplus$).} The cyan background depicts the background planetesimal densities (other planetesimals except for these protoplanets). The two gray vertical line marks the $r_0 \pm w_{\rm ring}$ width of the ring. The horizontal lines separate the three phases of the evolution, see main texts.}
\end{figure}

Different from pebble accretion in smooth discs, thanks to the highly concentrated dust inside the ring, the efficiency of pebble accretion is significantly enhanced. The enhancement is due to both the higher pebble density at the disc mid-plane and the lower headwind around the ring peak. The former condition is naturally met as the planetesimals can form only in the dense mid-plane with a high mass concentration inside the ring. Therefore, pebble accretion becomes efficient as soon as the planetesimal is formed! As \fg{timescale} shows, the growth timescale of the first planet is less than $0.1$~Myr once it was born. The blue and yellow lines in \fg{timescale} show the nominal pebble accretion timescale
\begin{equation}\label{eq:t_PA}
\begin{aligned}
    t_{\rm PA} = \frac{m}{\dot{M}_{\rm PA}}
    &= \left\{
    \begin{array}{lr}
        0.35\frac{m^{1/3}M_\star^{2/3}}{{\rm St}^{2/3}\Sigma_{\rm peb} r^2 \Omega_K}
         & \rm 2D\,regime\\
        0.20\frac{M_\star h_{\rm peb}}{{\rm St} \Sigma_{\rm peb} r^2 \Omega_K}
         & \rm 3D\,regime
    \end{array}
    \right.
\end{aligned}
\end{equation}
where we take $\Sigma_{\rm peb}$ the surface density at the ring peak and assume the relative velocity $\Delta v$ is determined by the Keplerian shear \eq{v_sh} and $f_{\rm set} = 1$.

\begin{figure}
    \centering
    \includegraphics[width=0.99\columnwidth]{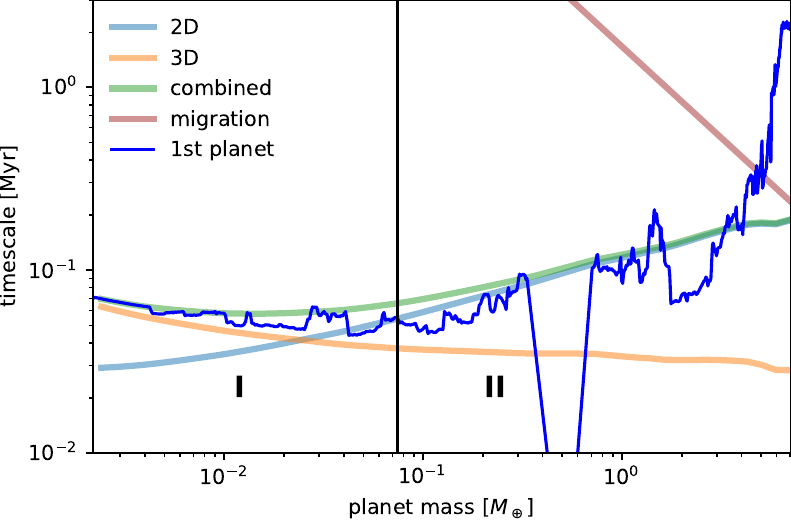}
    \caption{\label{fig:timescale} Pebble accretion and migration timescales as functions of protoplanet mass. The dark blue line represents the measured growth timescale $m/\dot{m}$ of the first planet that reaches the inner domain boundary in the \texttt{cr-tdefault} run. The light brown line indicates the type-I migration timescale to pass the ring $w_{\rm ring}/r_0 \times t_{\rm mg}$. The light blue, orange, and green lines indicate the analytical calculation of 2D, 3D, and combined pebble accretion timescale $m/\dot{M}_{\rm PA}$, assuming a surface density equal to that of at the ring peak (\eq{dotMPA} and \eq{RPA}). The ring mass is increasing while the first planet is growing, which initially decreases the pebble accretion growth timescale. The deep dip in the growth timescale around $1 M_\oplus$ records a collisional merger. The vertical black line marks the transition from phase I to phase II.}
\end{figure}

The evolution of protoplanets can be divided into three phases depending on the evolution of the pebble accretion rate and the ring mass (see \fg{stage}). In the first phase, the number of the planetesimal is still small. Pebble accretion has negligible influence on the evolution of the ring. Almost all incoming pebbles end up in the ring, thus the ring mass linearly grows with time at a slope that is the same as $\dot{M}_{\rm net}=\dot{M}_\mathrm{ext}-\dot{M}_\mathrm{leak}$. Meanwhile, planetesimals continue forming. Thus, the total pebble accretion rate monotonically increases with time because of both the elevated number of planetesimals and the growth of each planetesimal. Yet, as shown in the middle panel of \fg{stage}, the fraction of pebble accretion contributed by the largest particle keeps decreasing in phase I. The accretion rate of each planetesimal differs little since the pebble accretion is not a runaway mechanism \citep[][$\dot{M}_{\rm PA}\propto m$ in 3D regime, see also \eq{RPA}]{Ormel2017}.

\begin{figure}
    \centering
    \includegraphics[width=0.99\columnwidth]{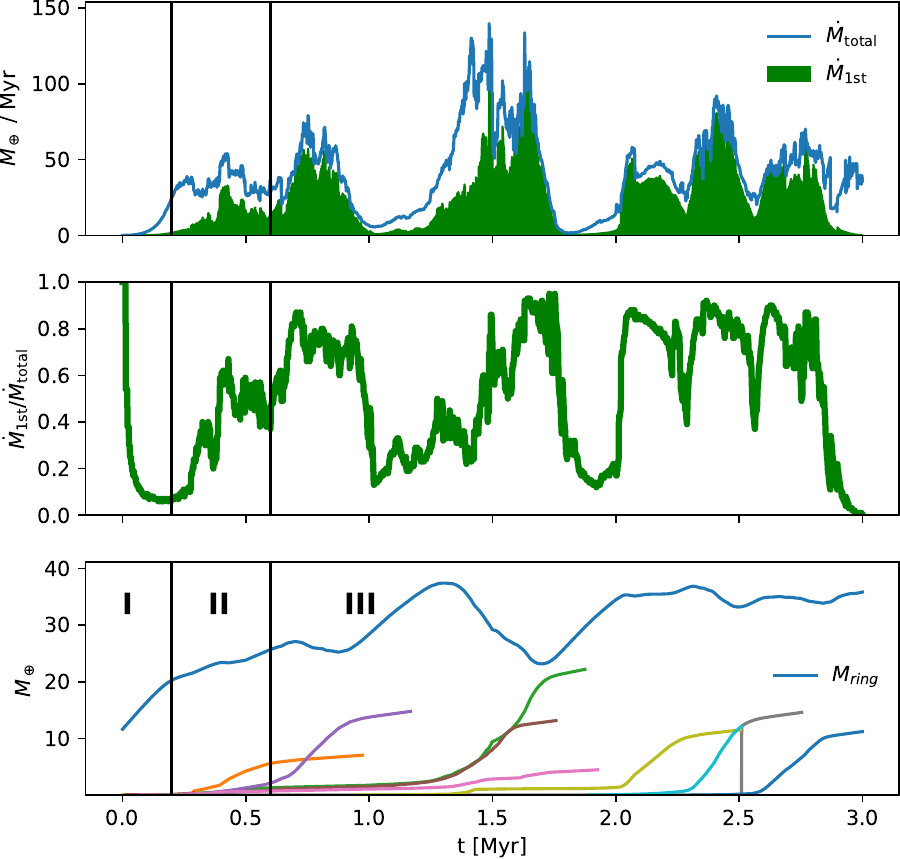}
    \caption{\label{fig:stage} \textbf{Top:} Total pebble accretion flux of all planetesimals (blue line). The green region shows the contribution from the most massive body still inside the simulation box (beyond $r_{\rm in}$). \textbf{Middle:} Fraction of the contribution from the most massive body over the total pebble accretion mass flux. \textbf{Bottom:} Mass of the ring (blue) and mass of the planets that have once been the most massive bodies (colours). When the planet migrates past the inner boundary of the simulation box, the line terminates. The vertical line at 2.5 Myr marks a collision between two planets. The evolution can be divided into three phases, see main texts.}
\end{figure}

Around 0.2~Myr, the total pebble accretion rate becomes non-negligible compared with the mass flux fed to the ring, which slows down the growth of the ring. As demonstrated in the bottom panel of \fg{stage}, the slope of the ring mass flattens. At the same time, the eccentricity and inclination of planetesimals are excited by viscous stirring \hjadd{as shown in \fg{ei_m}}. \hjch{Smaller planetesimals are more easily stirred up.}{Yet, due to dynamical friction, larger planetesimals have relatively lower eccentricities and inclinations whereas smaller planetesimals have higher velocity-dispersion. In addition, the tidal interactions (type I torque) damp the eccentricity and inclination more sufficiently as the planetesimals grow (\eq{t_ei}). By virtue of} the small scale-height of pebbles in the clumpy ring model, the high \hjadd{planetesimal} inclination will \hjch{lower}{reduce} the efficiency of pebble accretion \hjch{ dramatically}{(\eq{hpeb})}, suppressing further growth of these planetesimals by pebble accretion. Conversely, bigger planetesimals \hjch{tend to damp their}{with lower} inclination \hjrem{by planetesimal dynamical friction and} stay closer to the mid-plane region where the pebble density is the highest. As shown in \fg{timescale}, the first and also the biggest protoplanet transitions from the 3D accretion regime to the 2D accretion regime, in which the pebble accretion cross section is higher than the pebble scaleheight. The measured growth timescale matches the analytical formula very well because the planet is staying close to the ring centre and keeps a low inclination. As a result, the biggest protoplanet grows the fastest, while the small bodies are effectively left behind \citep{LevisonEtal2012}. As shown in \fg{stage}, at the top panel, the total pebble accretion rate does not increase significantly. The green region, which represents the accretion rate of the largest embryos, keeps rising. The fraction of the accretion rate of the largest embryos over the total accretion rate is plotted in the middle panel of \fg{stage}. At the beginning of phase II, it is less than 0.1 and becomes dominant as time evolves.

\begin{figure}
    \centering
    \includegraphics[width=0.99\columnwidth]{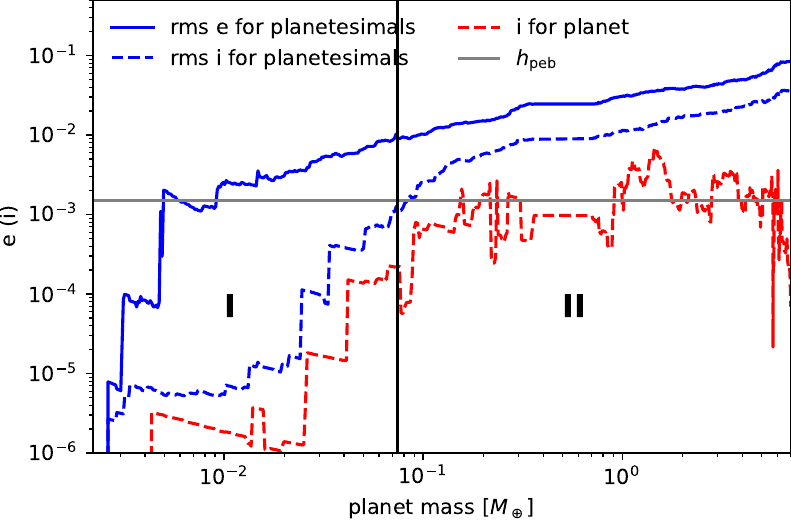}
    \caption{\label{fig:ei_m} \hjadd{Eccentricity (solid) and inclination (dashed) as functions of protoplanet mass in run \texttt{cr-default}. The red line represents the first planet and the blue lines indicate rms values of the planetesimals (other bodies except the most massive one). The gray solid line marks the inclination equal to the aspect ratio of the pebble disc $h_{\rm peb}$ at $r_0$. The vertical black line marks the transition from phase I to phase II.}}
\end{figure}

\hjch{In \fg{a_t_default}, we show the evolution trajectories of growing protoplanets. The colour of the trajectory line indicates the inclination of the embryos., the first planet keeps a low inclination, while the inclination and eccentricity of other planetesimals are excited. The cyan background corresponds to the density of planetesimals, in which the protoplanet (trajectory lines) are not included. the semimajor axes of planetesimals are expanding). Overall, in phase II, t}{The first planet keeps a low inclination in phase II, while the inclination and eccentricity of other planetesimals keep increasing (\fg{ei_m}, see also \fg{a_t_default} where the radial distribution of planetesimals are expanding).}
The biggest protoplanet stays closer to the mid-plane region and stirs up the newly formed planetesimals. The ring mass still grows but the growth rate becomes lower \hjadd{(middle panel of \fg{stage})}, since this protoplanet accretes from the ring significantly.

As the first planet has already grown to ${\sim}10\,M_\oplus$ at $t=0.6$~Myr, it feels sufficient gravitational torques from the gas disc and therefore leaves the ring via migration. After the planet migrates away from the massive ring, its accretion rate drops rapidly. Even though it can still accrete from the mass flux leaked from the ring, \hjrem{as discussed in previous literature,}the growth of the planet in a smooth pebble-flowing disc is slow and inefficient at such a large distance \citep[e.g.,][]{Ormel2017,LinEtal2018}. In \fg{mass_budget}, we stack the mass budget in different components as the function of time. The mass contribution accreted from the smooth disc is marked in red, which is \hjch{tiny}{insignificant} compared with those accreted from the ring. Meanwhile, as the first planet has left the ring, a smaller embryo within the ring will get the opportunity to grow. Since the external pebble flux that supplies the ring is fixed, the evolution becomes cyclical. In this cyclical phase III, the biggest planet inside the ring efficiently grows by pebble accretion and then migrates inward and leaves the ring behind. As shown in \fg{mass_budget}, regulated by the pebble accretion, the ring mass does not keep increasing but oscillates around a "saturated" value. The majority of the upcoming mass flux ends up in the largest embryos by pebble accretion from the ring. The long-lived ring acts as a planet factory where planets sequentially form and migrate towards the inner region of the disc. In the following section, we demonstrate the physical model to interpret the planet masses and the ring mass.

\begin{figure}
    \centering
    \includegraphics[width=0.99\columnwidth]{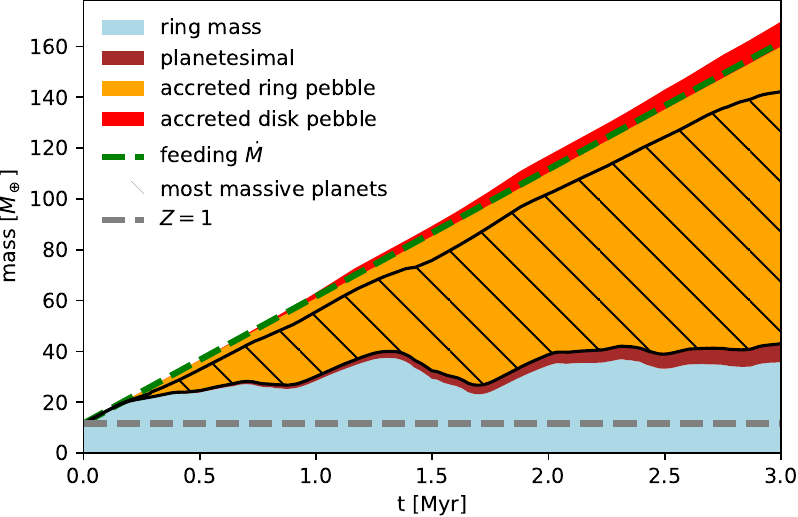}
    \caption{\label{fig:mass_budget} Evolution of the mass budget of different components in the default model. The blue region \hjch{shows}{indicates} the ring mass. The brown region shows the accumulated mass of planetesimals formed by streaming instabilities. The mass accreted to the planetesimals and protoplanets by pebble accretion from the ring (orange) and from the extended smooth disc component (red) are stacked on top. Dashed region marks the accumulated mass of planets that have ever been the largest planets inside the ring. Gray horizontal line marks the ring mass corresponding to $Z=1$. Green dashed line gives the accumulated mass fed to the ring, $\dot{M}_{\rm net}t$. The stacked value is a bit higher than $\dot{M}_{\rm net}t$ due to the accretion of the planet downstream.}
\end{figure}

\subsection{Mass of ring and planet in saturation phase}\label{sec:masses}
As we discussed in the default model, the ring mass gets into steady states after the first planet forms and migrates away from the ring, which is marked as phase III in \fg{stage}. We find this is a robust outcome for the vast majority of our simulations with type-I migration. We quantify the mass of the planets and the saturated ring in this section.

In most of our models, the mass of planetesimals formed directly by streaming instability is always small compared with the ring mass and the mass accreted to the planetesimals by pebble accretion (see \fg{mass_budget}). In other words, the mass flux fed into the ring is the same as the mass flux that planets are accreting from the ring. Therefore, the ring mass saturates at a value $M_{\rm ring,ss}$ at which the total accretion rate of planets is equal to the mass flux fed to the ring. The mass of the planet is then determined by the balance between planet accretion and migration, i.e, when the planet grows massive enough, it will migrate away from the ring centre and the accretion rate will rapidly drop because of the lack of feeding materials. By assuming there is always one planet with mass $m_{\rm rp}$ inside the ring, the typical accretion timescale of this planet is then $m_{\rm rp}/\dot{M}_{\rm net}$. And the timescale for this "representative" planet to migrate away from the ring is $\Delta t_\mathrm{ring} = 2 w_{\rm ring}/r_0 \times t_{\rm mg}(m=m_{\rm rp})$. By equating these two timescales, we obtain the representative planet mass
\begin{equation}\label{eq:m_rpr}
\begin{aligned}
    m_{\rm rp} &= 
    f_{\rm rp} \sqrt{\frac{\dot{M}_{\rm net}}{\Sigma_{\rm g}r_0^2\Omega_K} \frac{1}{f_{\rm mg}} \frac{w_{\rm ring}}{r_0}} h_{\rm g} M_\star  \\
    &= 18\,M_\oplus \times
    \left(\frac{f_{\rm rp}}{1.4}\right)
    \left(\frac{f_{\rm mg}}{1}\right)^{-\frac{1}{2}}
    \left(\frac{\dot{M}_{\rm net}}{50\,M_\oplus\,\rm Myr^{-1}}\right)^{\frac{1}{2}}
    \\
    &\times
    \left(\frac{\Sigma_{\rm g}}{3.6\,\rm g\,cm^{-2}}\right)^{-\frac{1}{2}}
    \left(\frac{w_{\rm ring}}{3.38\,\rm au}\right)^{\frac{1}{2}}
    \left(\frac{r_0}{74.2\,\rm au}\right)^{-\frac{3}{4}}
    \left(\frac{h_{\rm g}}{0.07}\right)
    \left(\frac{M_\star}{M_\odot}\right)^{\frac{3}{4}}
\end{aligned}
\end{equation}
where a fit constant $f_{\rm rp}=1.4$ is used.

As discussed above, when the ring mass saturates, the accretion rate of the planet is equal to the ring mass flux. It is shown in \fg{timescale} that the pebble accretion operates in the 2D regime for a planet with the mass of $m_{\rm rp}$. Therefore, by equating the 2D pebble accretion rate \eq{RPA} with $\dot{M}_{\rm net}$, we obtain a typical surface density
\begin{equation}\label{eq:Sig_pp}
\begin{aligned}
    \Sigma_{\rm tp} &=
    \frac{1}{2.9} \frac{\dot{M}_{\rm net}}{(St q_{\rm rp})^{\frac{2}{3}}\Omega_K r_0^2}
    = \frac{0.35 (f_{\rm mg}\Sigma_g)^{\frac{1}{3}} \dot{M}_{\rm net}^{\frac{2}{3}}}{(f_{\rm rp} h_{\rm g} \mathrm{St})^{\frac{2}{3}} (w_{\rm ring} G M_\star)^{\frac{1}{3}}} \\
    & = 0.13\,{\rm g\,cm^{-2}} \times
    \left(\frac{f_{\rm rp}}{1.4}\right)^{-\frac{2}{3}}
    \left(\frac{f_{\rm mg}}{1}\right)^{\frac{1}{3}}
    \left(\frac{\dot{M}_{\rm net}}{50\,M_\oplus\,\rm Myr^{-1}}\right)^{\frac{2}{3}}
    \\
    &\times
    \left(\frac{\Sigma_{\rm g}}{3.6\,\rm g\,cm^{-2}}\right)^{\frac{1}{3}}
    \left(\frac{w_{\rm ring}}{3.38\,\rm au}\right)^{-\frac{1}{3}}
    \left(\frac{St}{0.01}\right)^{-\frac{2}{3}}
    \left(\frac{h_{\rm g}}{0.07}\right)^{-\frac{2}{3}}
    \left(\frac{M_\star}{M_\odot}\right)^{-\frac{1}{3}}
\end{aligned}
\end{equation}
where we have substituted \eq{m_rpr} for $q_{\rm rp}=m_{\rm rp}/M_\star$ and $\Omega_K=\sqrt{GM_\star/r_0^3}$ in the first line. Substituting into \eq{M_ring}, this corresponds to a ring mass of
\begin{equation}\label{eq:M_ring_ss}
\begin{aligned}
    M_{\rm ring,ss} &= (2\pi)^{1.5} r_0 w_{\rm ring} f_{\rm rs}\Sigma_{\rm tp}
    = 5.5f_{\rm rs}  \frac{(f_{\rm mg}\Sigma_g)^{\frac{1}{3}} (w_r \dot{M}_{\rm net})^{\frac{2}{3}} r_0}{(f_{\rm rp} h_g \mathrm{St})^{\frac{2}{3}}(GM_\star)^{\frac{1}{3}}}
    \\
    &= 34\,M_\oplus \times
    \left(\frac{f_{\rm rs}}{1.8}\right)
    \left(\frac{f_{\rm rp}}{1.4}\right)^{-\frac{2}{3}}
    \left(\frac{f_{\rm mg}}{1}\right)^{\frac{1}{3}}
    \\
    &\times
    \left(\frac{\dot{M}_{\rm net}}{50\,M_\oplus\,\rm Myr^{-1}}\right)^{\frac{2}{3}}
    \left(\frac{\Sigma_{\rm g}}{3.6\,\rm g\,cm^{-2}}\right)^{\frac{1}{3}}
    \left(\frac{r_0}{74.2\,\rm au}\right)
    \\
    &\times
    \left(\frac{w_{\rm ring}}{3.38\,\rm au}\right)^{\frac{2}{3}}
    \left(\frac{h_{\rm g}}{0.07}\right)^{-\frac{2}{3}}
    \left(\frac{M_\star}{M_\odot}\right)^{-\frac{1}{3}}
    \left(\frac{\rm St}{0.01}\right)^{-\frac{2}{3}}
\end{aligned}
\end{equation}
where \hjch{$f_{\rm rs}=1.8$ is a numerical prefactor to correct the influence caused by the no-uniform surface density inside ring}{a numerical prefactor $f_{\rm rs}=1.8$ is taken to match the simulation results the best.} The predicted value $M_{\rm ring,ss}$ and the measured value $M_{\rm ring,ms}$ are listed in \tb{output} for every run. \Eq{M_ring_ss} approximates the measured ring mass very well. Substituting \eq{M_ring_ss} into \eq{m_rpr}, we \hjch{get}{obtain}
\begin{equation}\label{eq:Mp_Mring}
\begin{aligned}
    m_{\rm rp} &= 0.3 \times
    f_{\rm rs}^{-\frac{3}{4}}
    f_{\rm rp}^{\frac{3}{2}}
    f_{\rm mg}^{-\frac{3}{4}}
    \left(\frac{M_{\rm ring,ss}}{\Sigma_{\rm g}r_0^2}\right)^{\frac{3}{4}}
    {\rm St}^{\frac{1}{2}}
    h_{\rm g}^{\frac{3}{2}}
    M_\star
    \\
    &= 18 M_\oplus \times
    \left(\frac{f_{\rm rs}}{1.8}\right)^{-\frac{3}{4}}
    \left(\frac{f_{\rm rp}}{1.4}\right)^{\frac{3}{2}}
    \left(\frac{f_{\rm mg}}{1}\right)^{-\frac{3}{4}}
    \left(\frac{M_{\rm ring,ss}}{30 M_\oplus}\right)^{\frac{3}{4}}
    \\
    &\times
    \left(\frac{\Sigma_{\rm g}r_0^2}{0.002M_\odot}\right)^{\frac{3}{4}}
    \left(\frac{\rm St}{0.01}\right)^{\frac{1}{2}}
    \left(\frac{h_{\rm g}}{0.07}\right)^{\frac{3}{2}}
    \left(\frac{M_\star}{M_\odot}\right)
\end{aligned}
\end{equation}
which is independent of the unknown feeding mass flux. \hjch{Therefore, assuming that an ALMA ring is in steady state, if we can measure the mass of the ring, the Stokes number, the gas surface density and scaleheight, we can obtain the typical planetary core formed and delivered from the ring.}{Hence, if we know or can estimate the mass of the ring, the pebble aerodynamical properties, and the gas disc properties, \eq{Mp_Mring} provides the characteristic planet mass that emerges from the ring.}

\subsection{Similarities and differences between CR and PB models}\label{sec:SDCRPB}
Even though we focus on the CR model in our default run, the planet factory picture broadly holds for any ring that meets the following conditions:
\begin{enumerate}
    \item active planetesimal formation inside the pebble ring;
    \item pebble inside the ring have low approaching velocity towards these formed planetesimals;
    \item the ring is supplied by an external mass reservoir and therefore is long-lived;
    \item planets migrate inwards away from the ring.
\end{enumerate}
Conditions (i) and (ii) are required to warrant efficient pebble accretion. In other words, the criterion for planetesimal formation is consistent with massive volume pebble densities and the planetesimals can act as seeds of planet embryos. The second condition ensures pebble accretion falls in the settling regime, as otherwise (for strong headwinds) pebble accretion is severely suppressed \citep{VisserOrmel2016}. The longevity of the ring leaves enough time for planet growth; and, finally, the inward migration prevents a too massive planet to occupy the ring, which would otherwise destroy it, and allows a new embryo in the ring to grow (see \se{nmg} for further assessment).

\begin{figure}
    \centering
    \includegraphics[width=0.99\columnwidth]{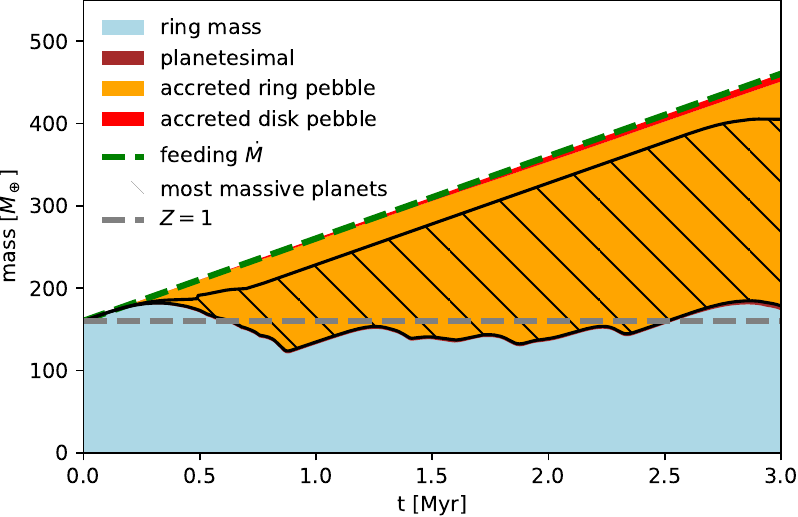}
    \caption{\label{fig:mass_budget_pb_default} Same as \fg{mass_budget} but for run \texttt{pb-default}. The evolution is similar between the clumpy ring and the pressure-bump-supported ring. A larger initial ring mass is required to reach $Z=1$ to start planetesimal formation. Yet, the mid-plane dust-to-gas ratio of the ring drops below unity during the evolution, hindering the planetesimal formation.}
\end{figure}

This picture applies to a planetesimal-forming ring held by a pressure bump as well. To illustrate this, in the \texttt{pb-default} run, we use the same parameters as the \texttt{cr-default} but assume that a permanent pressure bump supports the ring. The mass budget of run \texttt{pb-default} is shown in \fg{mass_budget_pb_default}. The ring initially grows when the planetesimals are still small when $t<0.3$~Myr. After that, most of the incoming mass flux ends up in the largest embryos by pebble accretion. The ring enters a steady state and the evolution is qualitatively the same as in the CR model.

An obvious difference between the CR model and the PB model is the ring mass. In order to reach the mid-plane dust-to-gas ratio $Z=1$, for pebbles with $\rm St = 0.01$, one magnitude higher initial ring mass is needed in the PB model compared with the CR model. Since the initial ring mass exceeds the predicted saturated ring mass, the planet grows so rapidly that the mid-plane dust-to-gas ratio drops below unity again, see  \fg{mass_budget_pb_default}. Planetesimal formation is, therefore, less efficient compared to the clumpy ring model (compare the "planetesimal" budget in \fg{mass_budget} and \fg{mass_budget_pb_default}). 
On the other hand, protoplanets migrate away from the ring when they reach the representative mass \eq{m_rpr}, at which point their accretion rate also exceeds the net feeding mass flux. The migration prevents the planet from consuming the ring, and therefore, the ring mass remains larger than the predicted ring mass, \eq{M_ring_ss}, during the evolution as marked in \tb{output}.

In the CR model, pebbles reside in a thin disc whose scaleheight is independent of St and $\delta_{\rm z}$. Thus, the loading timescale to reach $Z=1$, $t_\mathrm{load}$, chiefly depends on $\dot{M}_{\rm net}$. However, for PB-supported rings the pebble scaleheight, and therefore $t_\mathrm{load}$, are sensitive to St and $\delta_z$ (\eq{Hd}). For \texttt{pb-default} ($\rm St=0.01$ and $\rm \delta_{\rm z}=10^{-3}$), this wait amounts to $1.61$~Myr corresponding to a ring mass of $160 M_\oplus$. Discs with lower Stokes numbers (or higher $\delta_z$) require an even longer loading time, rendering the ring very massive but still sub-critical ($Z<1$). Planetesimal formation may never happen within the disc lifetime. In contrast, $t_\mathrm{load}$ is much shorter with a higher St or lower $\delta_z$ for which planetesimal formation can be triggered at lower surface densities (a less massive ring). For run \texttt{pb-St100} ($\rm St=0.1$) and \texttt{pb-a4} ($\rm \delta_{\rm z}=10^{-4}$), a $53 M_\oplus$ ring suffices to reach the dust-to-gas ratio $Z=1$. The lower initial ring mass means that fewer pebbles are available for planetesimals to accrete. Yet, the higher St and thinner pebble scaleheight enhance the pebble accretion efficiency (\eq{RPA}). Hence, the formation of the first planet is faster (shorter $t_{\rm 1}$ in \tb{output}) in runs \texttt{pb-St030}, \texttt{pb-St100} and \texttt{pb-a4}. Whereas at $\dot{M}_{\rm ext} = 100 M_\oplus\,\rm Myr^{-1}$, Stokes numbers ${>}0.02$ are catastrophic for clumpy rings due to the large leaking mass flux (see \se{dotM_St}), PB-supported rings can accommodate high St particles.

\subsection{Influence of type-I migration}\label{sec:nmg}
For both CR and PB models, type I migration plays a crucial role. Massive Planets migrate inward and eventually leave the ring, which terminates their rapid pebble accretion from the ring but promotes the longevity of the ring. To quantitatively understand the effect of migration on the evolution, we change the type-I migration prefactor $f_{\rm mg}$ (\eq{t_mg}) in the group of runs named with \texttt{mg}. We conduct runs with slower migration ($f_{\rm mg}=0.5$; \texttt{cr-05nmg}) and with faster migration ($f_{\rm mg}=2$; \texttt{cr-20nmg}). 

As expected from our physical model, the typical planet masses are higher (lower) in the slow (fast) migration run, and the saturated ring mass is lower (higher) accordingly, see \tb{output}. The outcome is intuitive. The migration speed decides the duration of the planet inside the ring, in which the majority of the core mass grows by the efficient pebble accretion. Thus, a slower migration speed allows more time for a single embryo to grow and consume the ring more, and vice versa.

In addition, in runs labeled \texttt{nmg} (no migration), we assume that the ring coincides with the zero-torque location. In this case, the planet formed inside the ring does not undergo type I migration, i.e., $f_{\rm mg}\to 0$. \hjrem{Such a situation can happen especially in the PB scenario, where the pressure bump arises from a local gas density variation and the co-rotation torque and Lindblad torque cancel each other (e.g., Masset et al. 2006).}The evolution of all \texttt{nmg} runs are similar. We show the evolution of the mass budget of the run \texttt{cr-nmg} in \fg{stage_nmg} as an example. For the first $0.6$~Myr, the evolution is similar to the run \texttt{cr-default}. This is because the planet mass in phase I and the beginning part of phase II are still small. However, after $0.6$~Myr, the evolution of the no migration runs is qualitatively different from the default runs since the most massive planet remains in the ring. Rather than leaving the ring, this planet keeps accreting and suppresses the growth of other embryos by scattering them away. Pebble accretion of the most massive planet becomes so dominant that it consumes the ring.

\begin{figure}
    \centering
    \includegraphics[width=0.99\columnwidth]{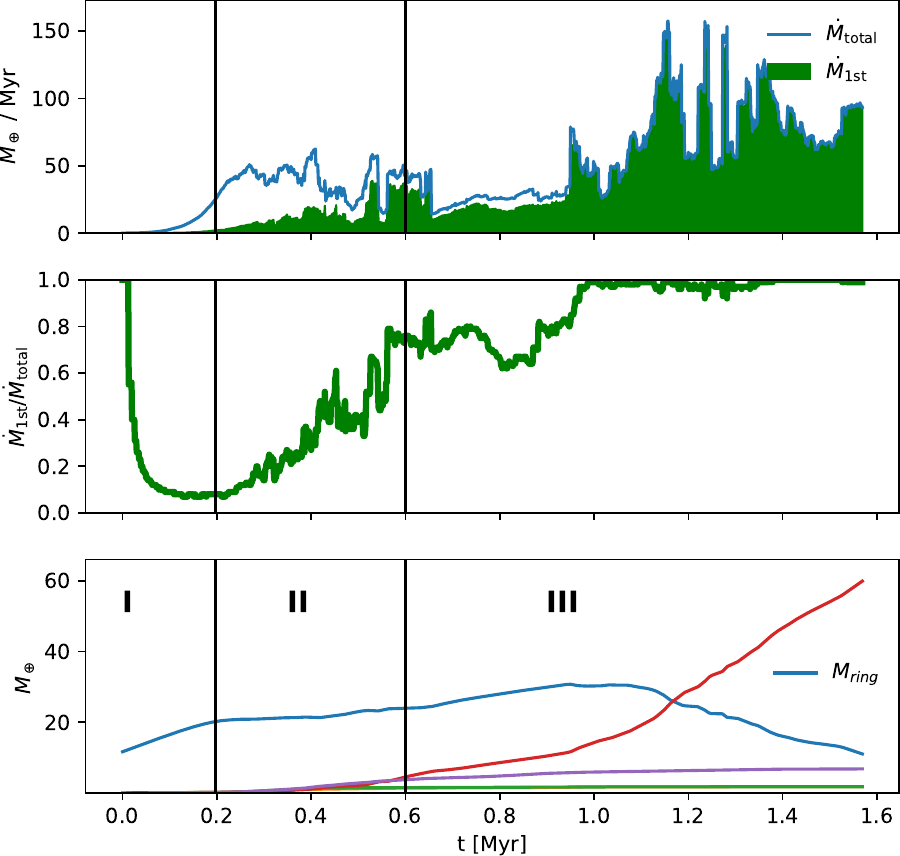}
    \caption{\label{fig:stage_nmg} Same as \fg{stage} but for the \texttt{cr-mng} run. The biggest planet dominates the pebble accretion and consumes the ring. }
\end{figure}

\subsection{\hjadd{Realistic pressure bump}}\label{sec:real_bump} 

\hjadd{In this section, we show the result of simulations where a pressure bump is the result of either a temperature bump or a gas density bump. Taking the locally isothermal disc assumption, the pressure follows
\begin{equation}
    p(r) = c_s(r) \Sigma_{\rm g}(r) \Omega_K(r) / \sqrt{2\pi}.
\end{equation}
Different from before, we change either $c_s(r)$ or $\Sigma_g(r)$ concomitant with $p(r)$ and also let these quantities affect the planet migration rate \eq{t_mg} and its prefactor $f_\mathrm{mg}=(2.7+1.1\beta)/3.9$ as discussed in \se{pl_mg}.}

\hjadd{We first consider the case that the pressure bump is purely caused by a local temperature bump in run \texttt{pb-t-bump}. We assume that a Gaussian bump of the sound speed
\begin{equation}
    c_{s, \rm Gss}(r) = \frac{\sqrt{2\pi}p_0}{\Sigma_{\rm g}(r)\Omega_K(r)} \exp\left(-\frac{(r-r_0)^2}{2w_{\rm pb}^2}\right)
\end{equation}
triggers the pressure bump. To minimise the influence of the temperature bump on the global disc, we relax the sound speed back to the smooth profile $c_s(r)$ when it is away from the ring centre. Thus, we set the sound speed
\begin{equation}\label{eq:c_s_tp}
    c_{s, \rm tp}(r) \equiv \sqrt{\frac{k_B T_{\rm tp}(r)}{\mu m_{\rm H}}} = f_{\rm bkg} c_s(r) + (1-f_{\rm bkg}) c_{s, \rm Gss}(r)
\end{equation}
in simulation \texttt{pb-t-bump}, where $f_{\rm bkg}$ is zero around the ring and unity away from the ring centre (see \eq{f_bkg} and \fg{f_bkg}). The radial profile of the gas surface density and temperature are shown in \fg{rad_prof_pb} in orange, as well as the corresponding pressure profile and $\eta$. Due to the temperature variation, the pebble scaleheight is slightly higher around the ring peak and lower interior to the ring peak, which have almost no influence on pebble accretion and planet migration. The outcome of run \texttt{pb-t-bump} is therefore almost the same as run \texttt{pb-default} as can be seen from the indicators listed in \tb{output}.}

\hjadd{Then, in run \texttt{pb-d-bump}, the pressure bump is caused by a local gas surface density variation, 
\begin{equation}\label{eq:Sig_gss}
    \Sigma_{\rm g, Gss}(r) = \frac{\sqrt{2\pi}p_0}{c_s(r)\Omega_K(r)} \exp\left(-\frac{(r-r_0)^2}{2w_{\rm pb}^2}\right),
\end{equation}
whereas the temperature profile remains smooth. Similar to \eq{c_s_tp}, we insert the density variation into the smooth gas disc $\Sigma_{\rm g}(r)$
\begin{equation}\label{eq:Sig_g_dp}
    \Sigma_{\rm g, dp}(r) = f_{\rm bkg} \Sigma_{\rm g}(r) + (1-f_{\rm bkg}) \Sigma_{\rm g, Gss}(r)
\end{equation}
which together with the corresponding profiles of pressure and $\eta$ are plotted in \fg{rad_prof_pb} in green. As shown in the bottom panel of \fg{rad_prof_pb}, the gas density bump qualitatively changes the type-I migration rate, as it leads to a zero-torque location at $\sim\!73$~au. This is the location where a planet is expected to be halted \citep[e.g.][]{MassetEtal2006}. As \fg{stage_d_bump} shows, a single planet grow rapidly from the ring and consume the ring, similar to the no-migration run \texttt{cr-nmg} and \texttt{pb-nmg}. The major reason is that the planet trapping location is close to the pebble ring peak, thus pebble accretion remains efficient for the planet trapped at that location.}

\begin{figure}
    \centering
    \includegraphics[width=0.99\columnwidth]{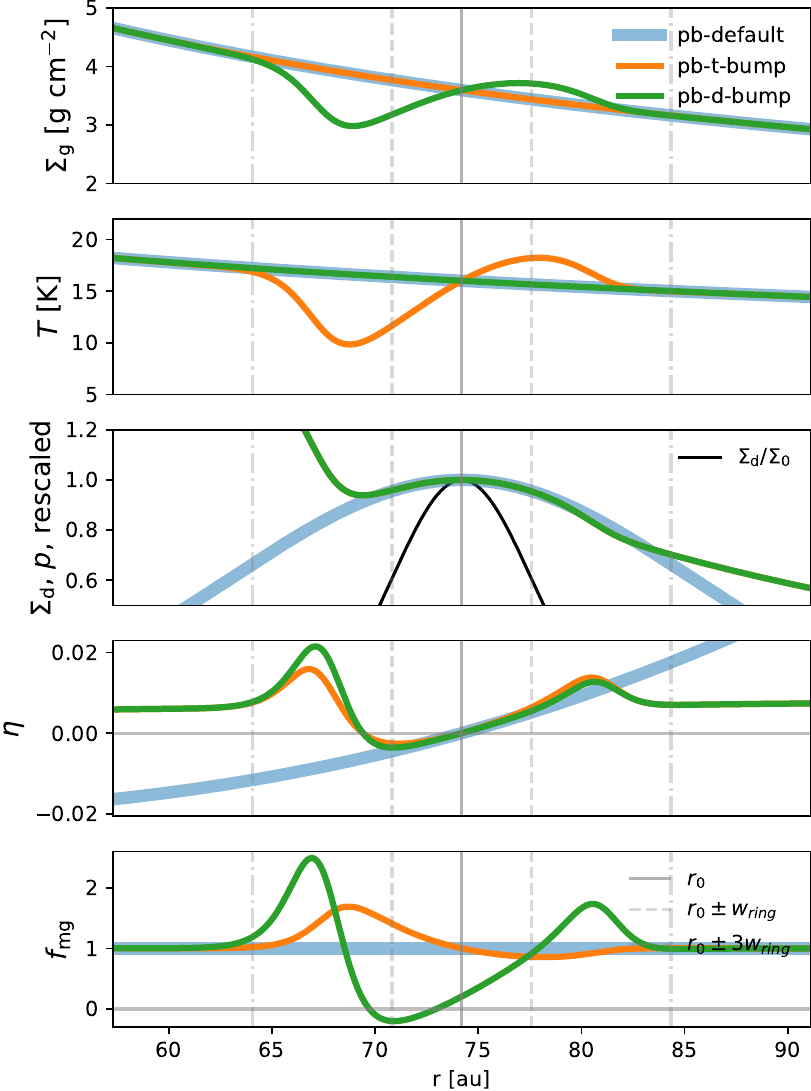}
    \caption{\label{fig:rad_prof_pb} \hjadd{From top to bottom, the five panels show the gas surface density profile, temperature profile, pressure profile, the dimensionless measure of the radial pressure gradient $\eta$, and the migration prefactor $f_\mathrm{mg}$ for runs \texttt{pb-default} (blue), \texttt{pb-t-bump} (orange), and \texttt{pb-d-bump} (green). The middle panel shows in addition the dust density profile (black curve).}}
\end{figure}
\begin{figure}
    \centering
    \includegraphics[width=0.99\columnwidth]{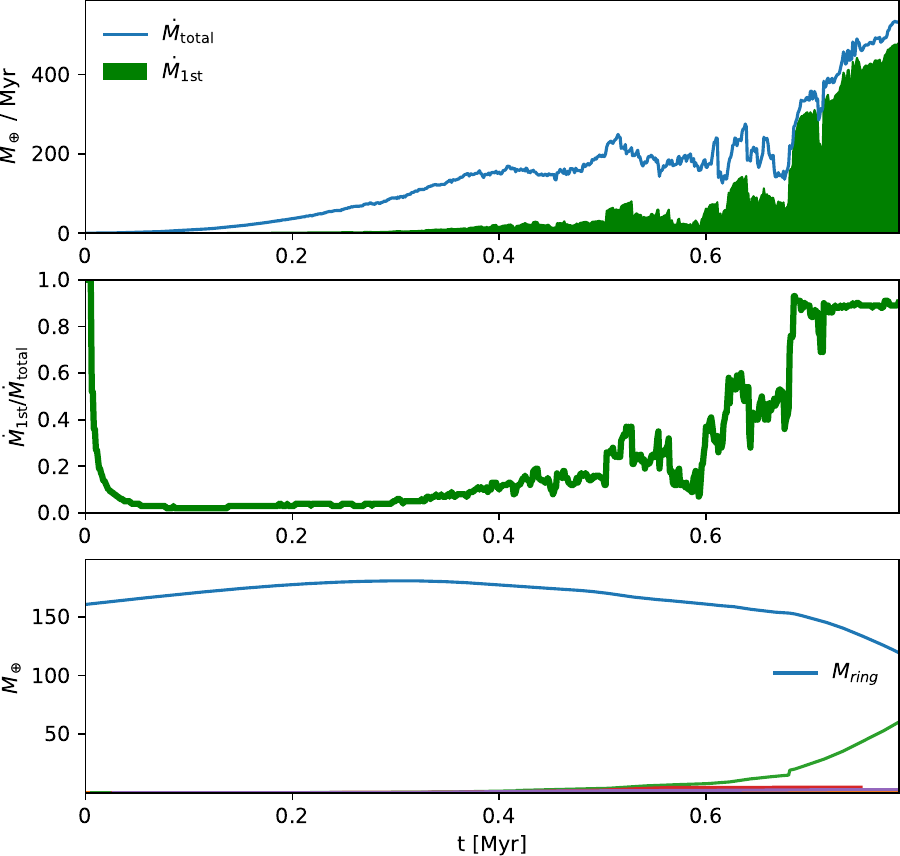}
    \caption{\label{fig:stage_d_bump} \hjadd{Same as \fg{stage} but for the \texttt{pb-d-bump} run, which features a migration trap. The biggest planet dominates the pebble accretion and consumes the ring, similar to other runs without migration.}}
\end{figure}

\hjadd{To better understand this result, we derive an expression for the trapping location, assuming that the surface density is dominated by the Gaussian-like component ($f_\mathrm{bkg}\approx0$). For locally isothermal discs, a trapping location ($f_\mathrm{mg}=0$) corresponds to \hjadd{a gas surface density slope} $\beta = -2.45$. In the case where $w_{\rm pb} \ll r_0$, by solving ${\rm d}\log{\Sigma_{\rm g, Gss}}/ {\rm d}\log{r} = -2.45$, the zero-torque location can be derived as $r_{\rm zt} = r_0 - 0.7 w_{\rm pb}^2/r_0$. The zero-torque location is always interior to the pebble ring peak $r_0$, and the offset reads
\begin{equation}
    \Delta r_{\rm zt} \equiv r_0 - r_{\rm zt}  = 0.7 \frac{w_{\rm pb}^2}{r_0} = 0.7 \frac{\delta_{\rm r}+{\rm St}}{\delta_{\rm r}}\frac{w_{\rm ring}^2}{r_0}
\end{equation}
where in the last step, we inserted the relation between the pressure bump width and the dust ring width (\eq{wring}). Therefore, for a fixed dust ring width the offset will be larger for larger Stokes number and weaker radial dust diffusivity parameter $\delta_{\rm r}$. In our default setup, we choose the ratio ${\rm St}/\delta_{\rm r} = 10$, following analysis of several DSHARP discs \citep{RosottiEtal2020}. For these values, the offset is only ${\sim}1$~au, much less than $w_\mathrm{ring}$. If we demand, on the other hand, that the zero-torque location is located significantly away from the dust ring, $\Delta r_\mathrm{zt} \gg w_\mathrm{ring}$, ${\rm St}/\delta_{\rm r}$ must be chosen much larger, perhaps by a factor of 10. In that case, it would render the pressure bump much wider ($w_\mathrm{pb}\gg w_\mathrm{ring}$) and very prominent indeed for $f_{\rm bkg} \to 0$ to hold in \eq{Sig_g_dp}. This situation is more reminiscent of an inner cavity rather than a local perturbation on top of a smooth disc structure. Hence, we conclude that for rings, fueled by local pressure bumps, the zero torque location cannot be significantly offset from the ring's centre.}

\hjadd{In addition, for the distances considered in our study (${\sim}75$~au), planets start migrating only after reaching $1 M_\oplus$ in locally isothermal smooth disc (See \fg{a_t_default}). So even if the zero-torque location was significantly away from the ring peak, rings can still produce planets up to Earth mass. In other words, the growth of the planets is already sufficient when they are still in the ring. This is different from, e.g., in \citet[][]{GuileraSandor2017}, where the embryos are not always initialized at the ring peak and can quickly migrate towards the zero-torque location as the migration timescales at ${\sim}5$ au are much shorter.}

\subsection{Parameter study}

\subsubsection{Initial mass of planetesimals}\label{sec:ini_mass} 
\begin{figure}
    \centering
    \includegraphics[width=0.99\columnwidth]{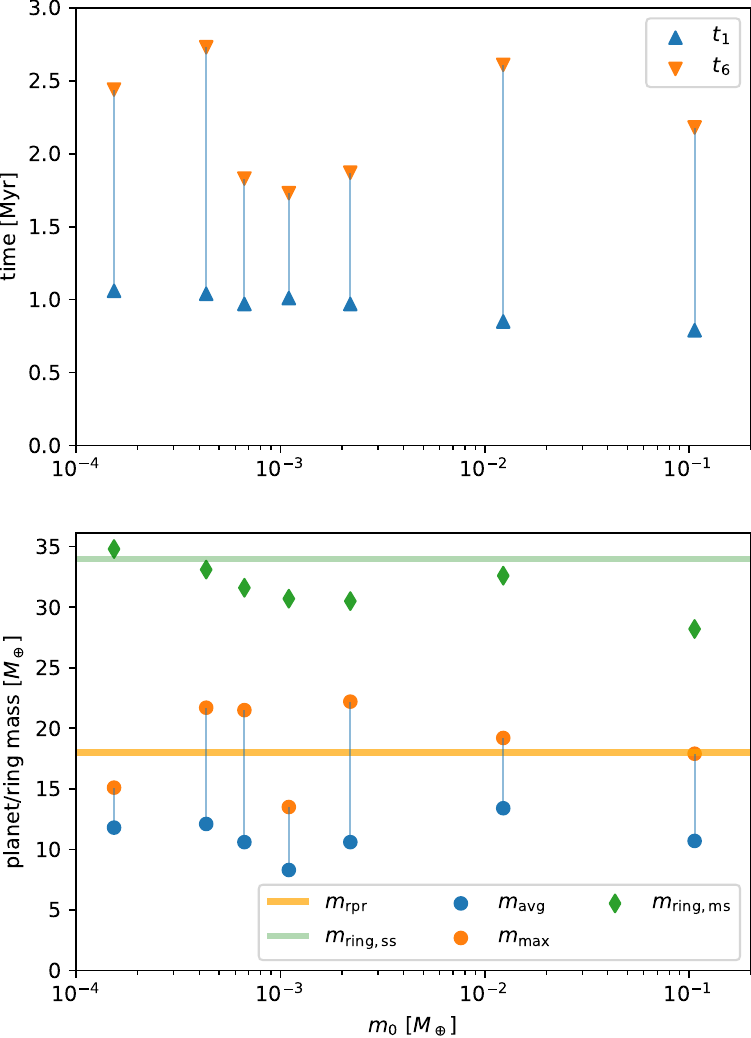}
    \caption{\label{fig:m0_tm} Output data of runs with different initial masses from Ceres mass to Mars mass from \tb{output}. \textbf{Top:} Time when the first (sixth) planet past $r_{\rm in}$, $t_1$ ($t_6$), is marked in blue (orange). \textbf{Bottom:} Dots mark the average and maximum mass of the six planets in each run when they pass $r_{\rm in}$. Green diamonds mark the measured ring masses between $t_1$ and $t_6$. Light orange and green lines indicate the predicted representative planet mass \eq{m_rpr} and the saturated ring mass \eq{M_ring_ss} respectively.}
\end{figure}

Since the mass of the planetesimals formed by SI is still under debate, we test a number of initial masses $m_{\rm 0}$ in our simulations, ranging from Ceres ($1.5\times10^{-4}M_\oplus$) to Mars ($0.1M_\oplus$), see \tb{input}. In \fg{m0_tm}, we visualize the growth time (upper panel) and the measured masses of planets and rings (lower panel) in runs with different initial masses. Interestingly, the simulation results almost show no dependence on the initial mass.

In contrast to pebble accretion in smooth discs, inside rings, the pebble density is very high and the relative velocity between pebbles and planetesimals is small. Therefore, planetesimals efficiently accrete pebbles once they are born, already at Ceres-masses. For our initial ring mass $M_{\rm ring,0} = 11.6 M_\oplus$, the e-folding timescale of pebble accretion in the 3D regime is less than $0.1\, \mathrm{Myr}$ and independent of the mass of the planetesimal (initial point in \fg{timescale}). For runs with an initial mass $m_{\rm 0} < M_{\rm Pluto}$, pebble accretion starts in the 3D regime. Due to the short e-folding growth timescale in the 3D regime, the duration of the planet in the 3D regime is short compared with the total growth timescale. In the 2D pebble accretion regime, the growth timescale increases with planet mass (\eq{t_PA}). Therefore, the majority of the growth time is spent in the final doubling of the planet's mass. Consequently, the time when the first planet passes the inner boundary after the first planetesimal formed, $t_{\rm 1}$ (blue triangles in \fg{m0_tm}), hardly depends on the initial planetesimal mass.

For run \texttt{cr-mMoon} and \texttt{cr-mMars} (the rightmost two columns in \fg{m0_tm}) the initial mass is so massive that pebble accretion starts in the 2D regime, during which the pebble accretion timescale is scaling with the mass of protoplanet. Thus $t_{\rm 1}$ shows a weak decreasing trend in these two columns.  On the other hand, for smaller planetesimals, the collision time-scale $t_{\rm coll} \sim \rho_\bullet R_{\rm plts} / \Sigma \Omega_K$ is shorter. Therefore, planetesimal mergers are more important in runs with smaller $m_{\rm 0}$ and speed up the growth. Finally, we also see some stochasticity due to the N-body interaction which results in a spread in quantities like $t_1$ and $t_6$. Altogether, the time to form 6 planets that can pass the inner boundary $t_{\rm 6}$ (orange triangles in \fg{m0_tm}) shows no dependence on initial mass at all.

In the lower panel of \fg{m0_tm}, we mark the measured average and maximum mass of the six planets that passed the inner boundary of the simulation domain with blue and orange dots, and the measured ring mass in green. These output masses of both rings and planets show no dependency on $m_{\rm 0}$ as well, consistent with our prediction in \eq{m_rpr} and \eq{M_ring_ss} (horizontal lines). In the saturation phase, the evolution is determined by the most massive planet, which always dominates the pebble accretion. And the mass of the most massive planet is self-regulated by migration. The initial planetesimal has a negligible influence on this self-regulation cycle.

Although we cannot conduct experiments with even smaller planetesimal masses due to computational constraints, the independence of the result on the initial planetesimal mass should hold also for lower masses. The reason that pebble accretion is so effective is that the eccentric velocity $e v_K \ll v_\ast$ \hjadd{(\eq{f_set})}. As shown in \fg{ei_t}, planetesimals are readily excited to eccentricities similar to the Hill value of the largest bodies $e\sim (q_{\rm 1st}/3)^{1/3}$ to enter the dispersion dominated regime (for which $i\sim e/2$) after $t=0.1\,\rm Myr$. In the dispersion dominated regime, however, the efficiency of the viscous stirring strongly decreases with eccentricity \citep[e.g.,][]{IdaMakino1993}. Balancing the viscous stirring timescale with the growth timescale (constant for pebble accretion in the 3D regime), an equilibrium eccentricity for the planetesimals follows, which is not much greater than the Hill eccentricity $e_{\rm H} \equiv v_{\rm H}/v_K=(q_{\rm 1st}/3)^{1/3}$. We then obtain that $e v_K \sim v_H < v_\ast$ and the eccentricity of the more massive protoplanets will still be lower due to dynamical friction. Viscous stirring, therefore, is rather ineffective. As all velocities scale the same with planet mass, the results are independent of the initial planetesimal mass.

\begin{figure}
    \centering
    \includegraphics[width=0.99\columnwidth]{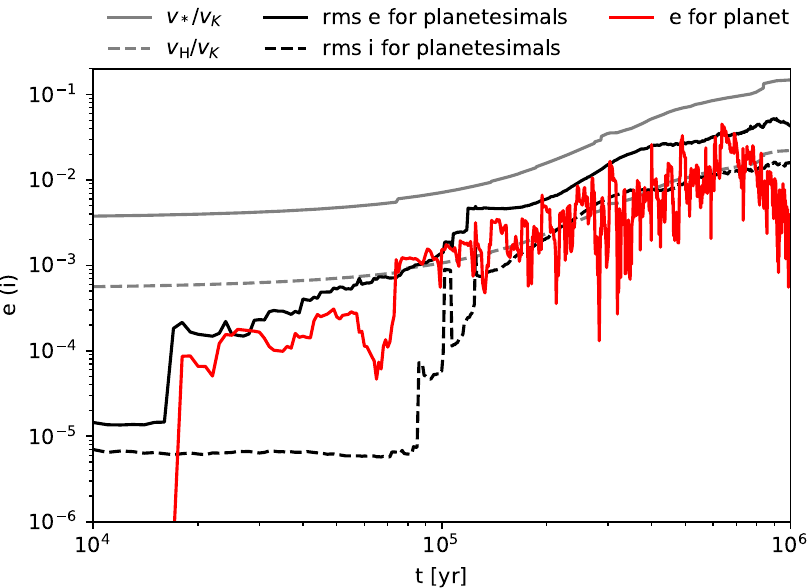}
    \caption{\label{fig:ei_t} Eccentricity (solid) and inclination (dashed) evolution in run \texttt{cr-mCeres}. The red line represent the most massive body and the black lines indicate rms values of the planetesimals (other bodies except the most massive one). The gray solid line marks the critical velocity threshold for pebble accretion $v_\ast$ and the gray dashed line marks the Hill eccentricity $e_{\rm H} \equiv v_{\rm H}/v_K$ of the largest body. Eccentricities follow $e_{\rm H}$ in the dispersion dominated regime and remain much lower than $v_\ast$.}
\end{figure}

\subsubsection{External mass flux and Stokes number}\label{sec:dotM_St}
In our simulations, we also vary the mass flux and Stokes number. Overall, the predicted and measured masses of planets and rings (column 6, 7, and 9, 10 of \tb{output}) show good correspondence, see \fg{ms_St} for a visualization.

\begin{figure}
    \centering
    \includegraphics[width=0.99\columnwidth]{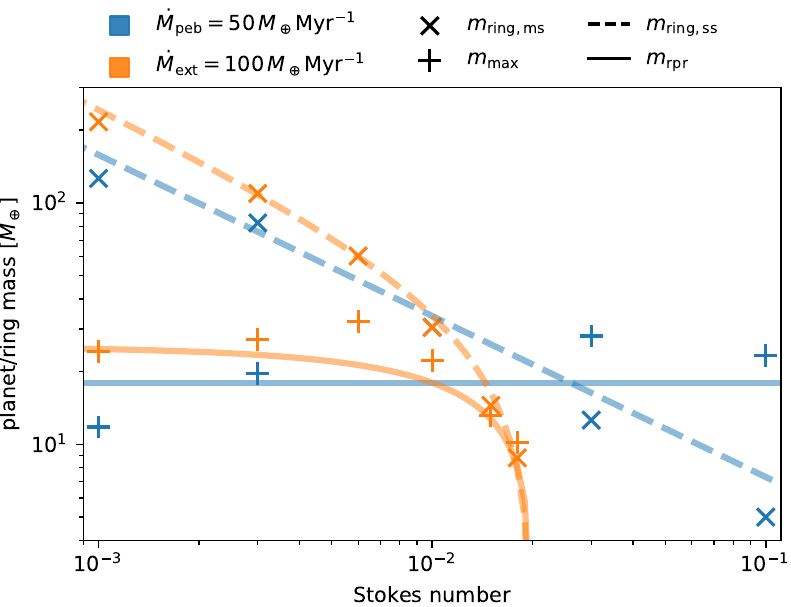}
    \caption{\label{fig:ms_St} Output data of runs with different Stokes number and external mass flux from \tb{output}. Blue markers belong to the run with the net mass flux feeding the ring $\dot{M}_{\rm net} = 50\,M_\oplus\rm\, Myr^{-1}$. Orange markers are runs with fixed external mass flux $\dot{M}_{\rm ext} = 100\,M_\oplus\rm\, Myr^{-1}$. Maximum planet masses $m_{\rm max}$ are marked as "\texttt{+}" and measured ring masses $m_{\rm ring,ms}$ are marked as "\texttt{x}". Solid and dash lines indicate the predicted representative planet mass \eq{m_rpr} and saturated ring mass \eq{M_ring_ss} respectively.}
\end{figure}

The dependence of the evolution on the external mass flux is intuitive, i.e., the higher the mass flux, the more available pebbles, the denser the ring, and the faster the growth of the planet.

The Stokes number, the aerodynamic size of pebbles, plays a crucial role in pebble accretion. In our model, there are three aspects controlled by the Stokes number:
\begin{enumerate}
    \item the settling of pebbles
    \item the leaking mass flux \eq{dotM_leak}
    \item the planetesimal formation rate \eq{dotM_plt}
\end{enumerate}
In our simulations, we always start with dust-to-gas ratio unity at the ring centre. For the CR model the Stokes number will not directly influence the initial ring mass, because the pebble scaleheight is set by the Kelvin--Helmholtz instability (\eq{Hc}). Independent of the assumption on the ring, for a given pebble surface density, the pebble accretion rates are higher for pebbles with large St in both 3D and 2D regimes, see \eq{RPA}. However, the leaking mass flux also increases with the Stokes number, because larger pebbles have a higher radial drift than smaller pebbles.

We investigate the role of St in CR rings by running two groups of simulations. First, in run \texttt{cr-St001} to \texttt{cr-St100} (blue markers in \fg{ms_St}), we 
keep the net feeding flux $\dot{M}_{\rm net}$ constant to isolate the influence caused by the leaking mass in our simulation. Since the leaking mass flux is linear with Stokes number in the CR model, we change the external mass flux accordingly to ensure that the net mass flux feeding the ring $\dot{M}_{\rm net} = \dot{M}_{\rm ext} - \dot{M}_{\rm leak}$ is always $50 M_\oplus\,\rm\,Myr^{-1}$. Therefore, the external mass flux is only $55 M_\oplus\,\rm Myr^{-1}$ for run with $\rm St=0.001$ while it increases to $550 M_\oplus\,\rm Myr^{-1}$ for run with $\rm St=0.1$.  Because of the low pebble accretion efficiency in the smaller Stokes case, the ring grows much more massive (blue dashed line in \fg{ms_St}), in line with our prediction in\eq{M_ring_ss}. From both the $t_{\rm 1}$ and $t_{\rm 6}$ indicators (see \tb{output}), it is clear that growth of a planet is slower in runs with lower St. However, as we predict in \eq{m_rpr}, there is no significant dependence between the planet mass and St (horizontal blue line in \fg{ms_St}). This is because massive planets migrate inward and leave the rapid-growth ring region. Termination of the growth depends on the migration behavior, which is determined by the gas properties of the disc rather than the pebble properties.

Then, in run \texttt{cr-St001F100} to \texttt{cr-St018F100}  (orange markers in \fg{ms_St}), we keep the external mass flux fixed at $100\,M_\oplus\,\rm Myr^{-1}$ and vary the Stokes number from $0.001$ to $0.018$. Now, both the feeding mass flux and the leaking mass flux change with Stokes number. When the Stokes number is low, the leaking mass flux is low as well and a larger portion of the external mass flux can be fed to the ring. Yet, the growth rate is still low because of the low pebble accretion efficiency. 
In contrast, when St is high, more pebbles will leak away from the ring to the inner disc. For pebbles whose Stokes number exceeds $0.02$, the leaking mass becomes larger than $100\,M_\oplus\,\rm Myr^{-1}$, which means that the clumpy ring will "vaporize" or never form in the first place. In \fg{ms_St}, the measured planet and ring masses rapidly drop when the Stokes number approaches $0.02$, as predicted by the orange lines. Therefore, a larger Stokes number is not necessarily beneficial to planet formation. For a fixed external mass flux of $100\, M_\oplus\,\rm Myr^{-1}$, the growth rate peaks for a Stokes number around $\rm St=0.01$, which shows smallest $t_{\rm 1}$ among all $\dot{M}_{\rm ext}=100\, M_\oplus\,\rm Myr^{-1}$ runs (see \tb{output}).

The Stokes number in our model also influences the planetesimal formation rate since the planetesimal formation timescale scales with the settling time scale, see \eq{dotM_plt}. Thus, pebble rings with larger Stokes numbers can form planetesimals faster. For the default value of the planetesimal formation efficiency parameter ($\zeta=10^{-3}$), pebble accretion and leaking always dominate the net mass flux fed to the ring while the flux represented by planetesimal formation is minor.

\subsubsection{Planetesimal formation efficiency}\label{sec:epsilon}

\begin{figure}
    \centering
    \includegraphics[width=0.99\columnwidth]{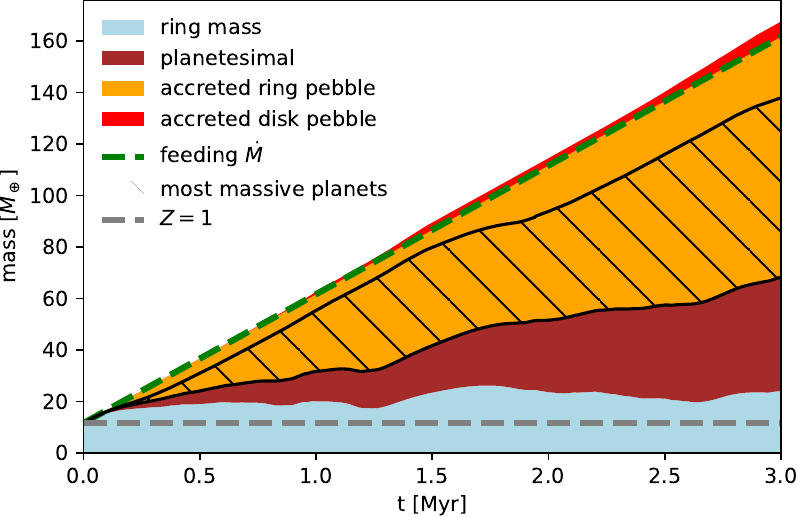}
    \caption{\label{fig:mass_eps2} Same as \fg{mass_budget} but for run \texttt{cr-e2}. The large amount of direct planetesimal formation (brown) limits the mass of the pebble ring significantly.}
\end{figure}

We investigate the planetesimal formation efficiency by varying the prefactor $\zeta$ in \eq{dotM_plt}. We test a higher $\zeta=0.01$ in run \texttt{cr-e2} and several lower $\zeta$ runs. As we show in \fg{mass_budget}, in the default model where $St=0.01$, the fraction of pebbles that collapse into planetesimals (by SI rather than ending up in planets by pebble accretion) is small. Therefore, the mass loss of the ring due to planetesimal formation is negligible compared with pebble accretion. This no longer holds for the run with higher $\zeta$, in which case planetesimal formation will significantly 
limit the pebble ring mass as shown in \fg{mass_eps2}. As a result, the measured saturated ring mass in run \texttt{cr-e2} is only $22.1 M_\oplus$, significantly lower than the predicted value $34\, M_\oplus$ in \eq{M_ring_ss}. Since planetesimal mergers is rare at these large radii, the large planetesimal budget does not help with the growth of the planet. The growth time of the first planet $t_{\rm 1}$ is similar between \texttt{cr-e2} and \texttt{cr-default}. Conversely, since the saturated ring mass is lower in phase III, growth of later forming planets requires more time. The average interval time $\langle\Delta\,t \rangle$ on which the ring spawns planets doubles in the higher $\zeta$ run (see \tb{output}). In the same amount of time, more smaller planetesimals are born but fewer big planets form. In particular, we note that further increasing the planetesimal formation efficiency is catastrophic for the survival of the clumpy ring. In JO21, even without accounting for pebble accretion, rings cannot survive in runs where $\zeta\gtrsim0.1$, since the rapid planetesimal formation drives the ring mid-plane dust-to-gas density ratio below unity. See \se{caveats} for further discussion.

On the other hand, for lower planetesimal formation efficiency (\texttt{cr-e4}, \texttt{cr-e5}), there is almost no difference with the default model. Ring mass removal is dominated by pebble accretion. Therefore, the planetesimal formation has a negligible impact on the ring's evolution.

\subsubsection{Location of the ring}\label{sec:r0}
Finally, we investigate how the results depend on the location of the ring. Previous literature has highlighted the influence of the disc radius on the planet formation \citep[e.g.][]{VisserOrmel2016,BitschEtal2019,Morbidelli2020,Chambers2021,VoelkelEtal2021i,JangEtal2022}. We setup a ring at $30$~au in run \texttt{cr-r30} and at $154$~au in run \texttt{cr-r154}, where the leaking mass flux change to $40\, M_\oplus\,\rm Myr^{-1}$ and $60 M_\oplus\,\rm Myr^{-1}$ respectively (\eq{dotM_leak}). Other disc parameters are all the same as the \texttt{cr-default} run.

Interestingly, both the planet mass and planet growth timescale show little dependence on the location of the ring (see \tb{output}), \hjch{in contrast to Morbidelli (2020) who report a significantly efficient growth when the ring is closer to the host star}{distinguishing from pebble accretion in smooth discs where planet growth is significantly enhanced when closer to the star \citep[e.g.,][]{IdaEtal2016,Ormel2017}}. One of the reasons for the insensitivity on disc radius is that the saturated ring mass, as derived in \eq{M_ring_ss}, scales with the orbital radius.
In the saturated phase, the net pebble flux fed to the ring will eventually be accreted by the planet. Since the planet can accrete the ring more efficiently at closer orbits, pebbles remain a shorter time in the ring.
Therefore, in the inner disc, the ring is less massive than the ring in the outer region, if we assume the same ring width. And the pebble accretion rate of the planet always reaches the net mass flux $\dot{M}_{\rm net}$ at the end. On the other hand, the timescale required to migrate through the ring width $\Delta t_\mathrm{ring}$ is independent of radius for our choice of the disc and ring model parameters, leaving a similar time for the planet to accrete pebbles from the ring. Therefore, as we derived in \eq{m_rpr}, the planet mass always approximates $\dot{M}_{\rm net} \Delta t_\mathrm{ring}$. We find the growth of planets in the clumpy ring model shows only a weak preference for lower radii.

\subsection{Post-ring evolution of planets}{\label{sec:post_ring}}
In the CR model, planets that have migrated and left the ring can still accrete pebbles downstream by virtue of the leaking mass flux.
Compared with pebble accreted from the ring, the fraction of pebbles accreted from the leaking mass flux is nevertheless small around the ring region (the red region above the green dash line in \fg{mass_budget}). We briefly discuss the subsequent evolution of planets after they have left the simulation domain.

\begin{figure}
    \centering
    \includegraphics[width=0.99\columnwidth]{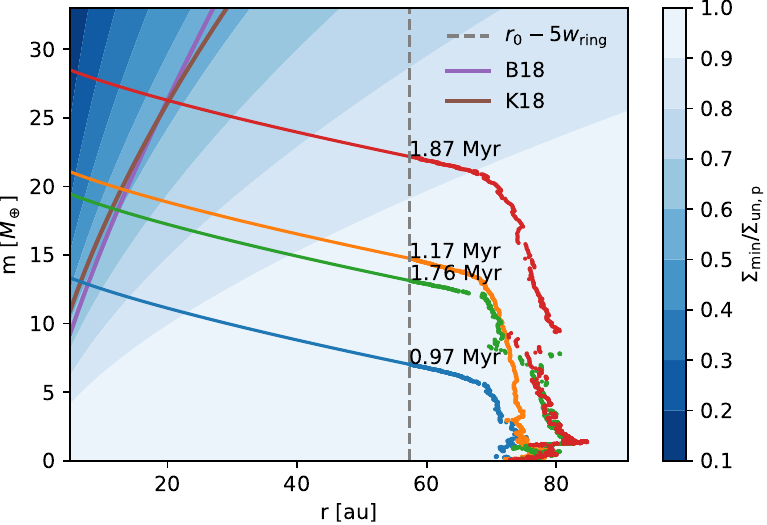}
    \caption{\label{fig:track} Trajectories of the first four planets larger than $5M_\oplus$ that passed the inner simulation domain in \texttt{cr-default}. Dots mark the evolution tracks of planets when they are still inside the N-body simulation domain. Solid lines in the same colours indicate the predicted trajectories following \eq{m_r_in}. Purple line "B18" marks the pebble-isolation mass from \citet{BitschEtal2018}. Brown line "K18" marks the gap opening mass from \citet{KanagawaEtal2018}. The time when planets pass the inner boundary $r_{\rm in}$ is labeled accordingly. The contours in the background show the expected gap depth in the gas disc carved by the planet.}
\end{figure}

We show the evolution tracks of the first four planets larger than $5M_\oplus$ that passed the simulation domain in \texttt{cr-default} in \fg{track}. The dots record the evolution track when these planets are still in the N-body simulation. Solid lines indicate their corresponding predicted trajectories based on \eq{m_r_in} (see \App{mr_tracer} for the mathematical expressions). The blue contours represent the expected gap depth carved by the planet at corresponding radii \citep{KanagawaEtal2018}
\begin{equation}
    \frac{\Sigma_{\rm min}}{\Sigma_{\rm un,p}} = \frac{1}{1+0.04K}
\end{equation}
where
\begin{equation}
    K = \left(\frac{m_{\rm p}}{M_\star}\right)^2\left(\frac{h_{\rm g}}{a_{\rm p}}\right)^{-5}\alpha^{-1}
\end{equation}
These planets keep growing by pebble accretion and migrating inward until they reach the pebble isolation mass, at which the gas gap carved by the planet is deep enough to halt the pebble flux at the outer edge and stop the pebble accretion \citep[e.g.,][]{LambrechtsJohansen2014,BitschEtal2018}. \footnote{\hjadd{Recently, \citet{SandorRegaly2021} showed that a planetary core can grow larger than the nominal pebble isolation when it is located inside a global pressure maximum.}} In \fg{track}, we plot the pebble isolation mass as the purple line according to Eq.~(11) in \citet{BitschEtal2018}. Planets that reach the pebble isolation mass then undergo runaway gas accretion and grow to gas giants \citep[e.g.][]{BitschEtal2019}. Meanwhile, the migration speed also slows down as the surface density of the bottom of the gap decreases \citep{KanagawaEtal2018}. The brown line indicates the gap opening mass \begin{equation}\label{eq:M_gap}
    M_{\rm gap} = 60\left(\frac{\alpha}{10^{-3}}\right)^{0.5}\left(\frac{h_{\rm g}}{0.07}\right)^{2.5} M_\oplus
\end{equation}
\citep[Eq.~(23) in ][]{KanagawaEtal2018}, which marks a transition from type-I migration to so-called type-II migration.

In addition, we test a lower disc turbulence ($\delta_{\rm z} = 10^{-4}$) in run \texttt{cr-a4}, in which planets grow faster and larger (see \tb{output}). This outcome is not because of the planet accretion inside the ring, since the pebble scaleheight of the clumpy ring is independent of $\delta_{\rm z}$. Yet, the low disc turbulence does help with the post-ring evolution. When the turbulence is suppressed, the disc pebble scaleheight is smaller. Therefore, the pebble accretion downstream is significantly enhanced. The largest planet reaches a mass of $50.7\,M_\oplus$ in \texttt{cr-a4} when it arrived at the inner boundary of the N-body simulation domain. At this mass, the planet may already have reached the condition for gap opening if we assume that the low dust vertical turbulence is indeed caused by the low gas viscosity (although $\alpha$ and $\delta_{\rm z}$ can be different as we mentioned below \eq{Hd}). The pebble isolation mass and gap opening mass are only 20~$M_\oplus$ at the ring location with $\alpha = 10^{-4}$. Therefore, this planet should have opened a gap in the disc already and started gas accretion.

\subsection{Planetesimal belt}\label{sec:plt_belt}

\begin{figure*}
\centering
\subfiguretopcaptrue
\mbox{}\hfill
\subfigure[$t = 0.5$ Myr]{\label{fig:mass_plt_a}\includegraphics[width=0.32\textwidth]{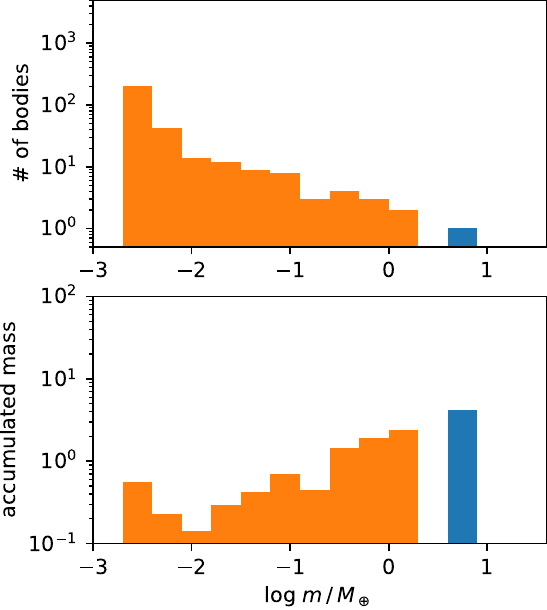}}\hfill
\subfigure[$t = 1.5$ Myr]{\label{fig:mass_plt_b}\includegraphics[width=0.32\textwidth]{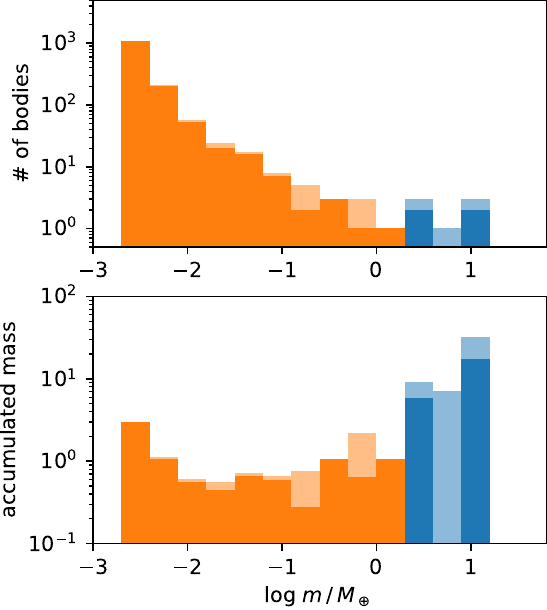}}\hfill
\subfigure[$t = 2.5$ Myr]{\label{fig:mass_plt_c}\includegraphics[width=0.32\textwidth]{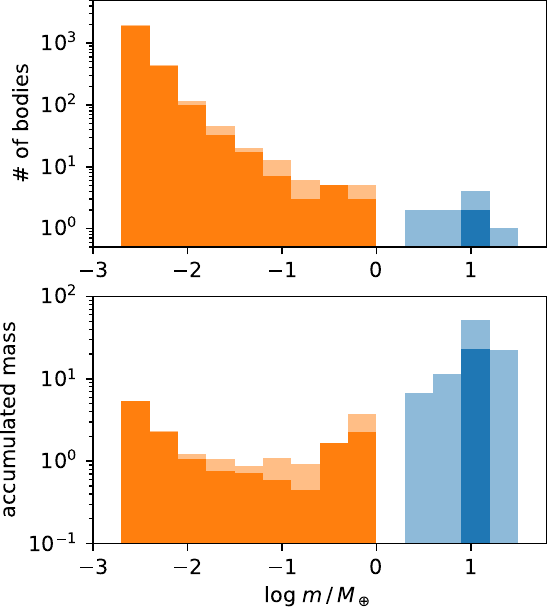}}\hfill
\hfill\mbox{}
\subfiguretopcapfalse
\caption{Mass distribution of bodies at $t=0.5$, 1.5, 2.5 Myr in \texttt{cr-default} run. Top row shows the histogram in terms of the number of bodies, and the bottom row shows the histogram in terms of accumulated mass in each bin. Blue bins mark the bodies classified as planets in our simulation (mass larger than $m_{\rm crit}$ and used for statistics in \tb{output}), all planets eventually migrate through the inner boundary of the simulation domain. Others are classified as planetesimals and coloured in orange. For both planet and planetesimals, the dark part remains in the N-body simulation domain, and the light part has already passed the inner boundary. In \texttt{cr-default}, all planetesimals start with Pluto mass, which is the cutoff on the left of each panel.}
\label{fig:mass_plt}
\end{figure*}

As proposed in JO21, the large amount of planetesimals formed in the pebble ring may evolve into a debris disc with the dispersal of the gas disc. In JO21, we neither considered any interaction between pebbles and planetesimal nor accounted for the dynamics of planetesimals. To balance the external pebble flux and the ring mass loss, the planetesimal formation efficiency is set to be $\zeta = 0.01$ in the default run, which results in a total mass up to $100 M_\oplus$ of planetesimals over a time-span of $5$~Myr.

In \fg{mass_plt}, we show the mass distribution of the bodies in \texttt{cr-default} at different times. We colour the planet population, whose $t_{\rm mg}(r=r_0)\leq 1 \rm Myr$, in blue and planetesimals , whose $t_{\rm mg}(r=r_0)> 1 \rm Myr$, in orange as we defined in \se{default_model}. The majority of the ring mass ends up in planets, while most of the planetesimals remain of low mass. The shallow region in the histogram are those bodies that have passed the inner simulation domain boundary. All planets migrate inward and leave the ring region eventually as shown in \fg{a_t_default}. Due to their low mass, planetesimals will stay around the ring location.

To understand the evolution of the planetesimal belt, we show the time-averaged "radial profile" of planetesimals' number density during [0,1], [1,2], [2,3]~Myr in \fg{n_plt_t}. As time evolves, the radial profile of the planetesimal belt becomes asymmetrical, characterized by a steep rise and a gradual decline. Planetesimals experience both resonant shepherding and outward dynamical scattering by the inward-migrating planets \citep[e.g.,][]{BatyginLaughlin2015}. As time evolves, planetesimals interior to the ring can be shepherded by migrating planets through mean resonance \citep[e.g.,][]{ShibataIkoma2019} and leave the inner boundary of the simulation domain. In contrast, planetesimals are more likely scattered outward by the inward-migrating planets and stay exterior to the ring, leading to a sharper inner edge and a flatter outer edge.

\begin{figure}
    \centering
    \includegraphics[width=0.99\columnwidth]{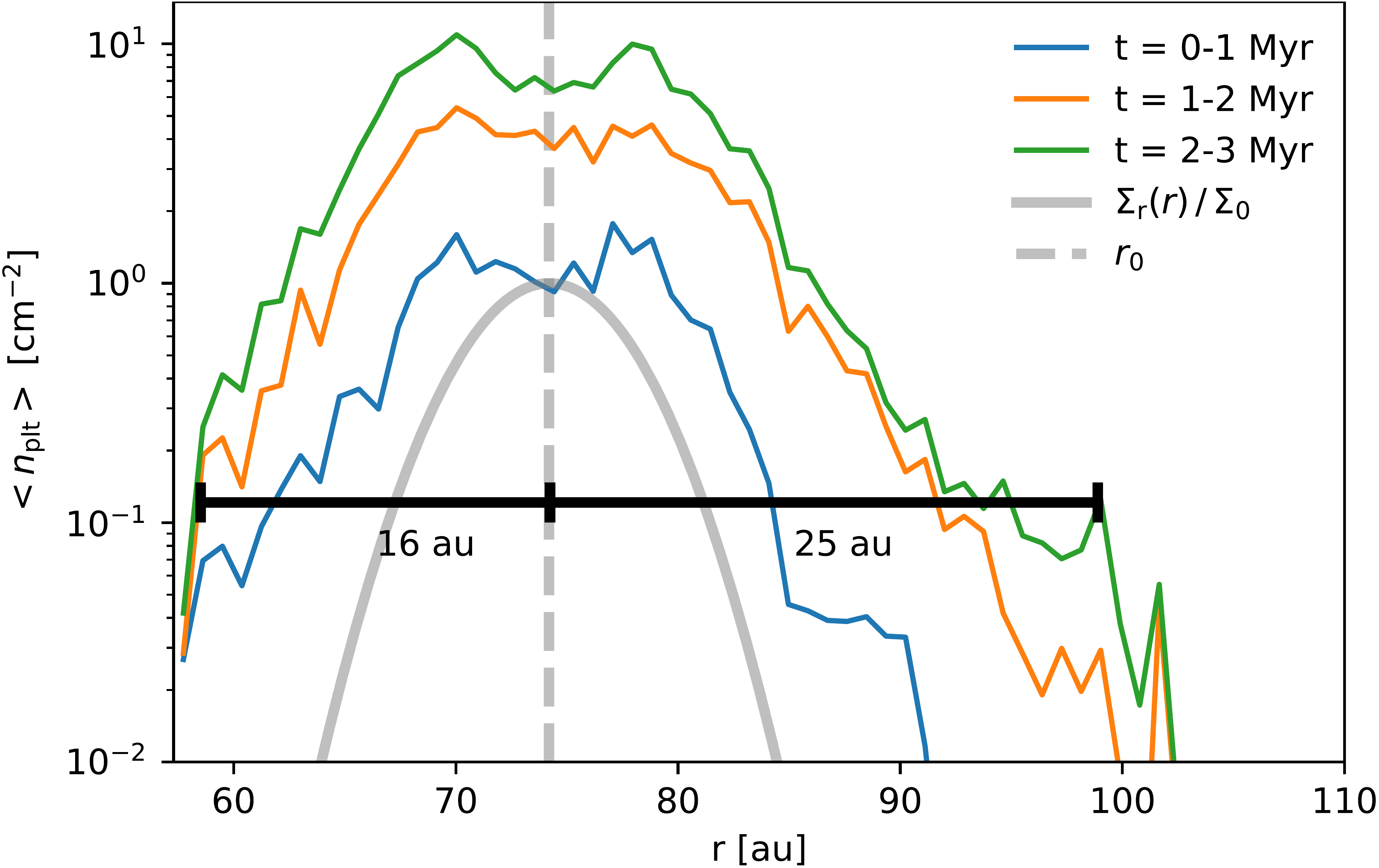}
    \caption{\label{fig:n_plt_t} The time-averaged planetesimal number density radial profile during 0--1, 1--2, and 2--3~Myr in \texttt{cr-default} run. Radial profile of the pebble ring profile is shown in gray. As time evolves, more planetesimals are scattered outward by the planet that migrated interior to the ring. Horizontal segments mark the $3\sigma$ width from the pebble ring centre by assuming half Gaussian distributions for two edges of the planetesimal belt respectfully. The outer edge is shallower than the inner.
}
\end{figure}

\section{Discussion}\label{sec:discussion}

\subsection{Caveats}\label{sec:caveats}
The model presented in this work employed a number of assumptions and simplifications.

First, in our simulation, \hjch{we aim to study the local evolution around the ring region}{we fix the pebble flux and keep it at a relatively high level of ${\sim}100 M_\oplus\,\rm Myr^{-1}$ for several million years and ignore the global evolution of the disc i}n order to isolate the effect caused by the uncertain disc evolution \citep[See][and references therein]{MiotelloEtal2022}. 
\hjadd{These high mass fluxes require a massive disc.} \hjadd{According to the \texttt{pebble-predictor} of \citet[][]{DrazkowskaEtal2021}, a disc with an initial mass of $1\,000 M_\oplus$ in solids and an characteristic radius of $300$~au can maintain a pebble flux $\dot{M}_{\rm ext} > 100 M_\oplus\,\rm Myr^{-1}$ for $\sim2$~Myr. A massive disc dust budget of $1000 M_\oplus$ is indeed high. Yet recent planet formation studies involving pebble accretion considered similarly abundant pebbles mass reservoirs \citep[e.g.,][]{IdaEtal2016,LambrechtsEtal2019,BitschEtal2019}. In addition, the current sample of very high resolution measurements in the mm continuum or scattered light is biased towards larger and brighter, and therefore more massive discs \citep[e.g.,][]{Andrews2020}. Moreover, even though many Class II discs have low dust masses \citep[e.g.,][]{AndrewsEtal2013,AnsdellEtal2016,AnsdellEtal2017,TripathiEtal2017}, discs may initially be much more massive. For example, \citet[][]{Xu2022} tests a parameterized model on the VANDAM survey and suggests that the majority of Class 0/I Discs in Orion are gravitationally unstable and as massive as the host star. Finally, recent studies adopting more robust methods of mass estimation also suggest that the disc gas mass –and by extension the dust mass– is higher than what is inferred from CO observations \citep[e.g.,][]{PowellEtal2017,PowellEtal2019,AndersonEtal2022,SturmEtal2022}.}

In reality, the pebble flux \hjch{could}{is also likely to} decay with time \citep[e.g.,][]{LambrechtsJohansen2014,VenturiniEtal2020,DrazkowskaEtal2021}. For discs where the dispersal timescale exceeds ${\sim}\rm Myr$ \hjadd{(the typical loading timescale for the ring, and the interval timescale of planet spawning, see \tb{output})}, the steady state solution we derived is valid. However, if the pebble flux decreases on a timescales shorter than ${\sim}\rm Myr$, the ring mass and the ensuing planet masses will decrease accordingly (\eq{m_rpr} and \eq{M_ring_ss}).  In future studies, \hjadd{it is worthwhile to consider} a \hjrem{more complete} model \hjch{of disc evolution}{accounting for time-dependent surface density and pebble flux,} \hjrem{pebble drift,} coagulation, and fragmentation \hjrem{needs to be considered} \citep[e.g,][]{GuileraEtal2020,VenturiniEtal2020,SchneiderBitsch2021i}.

The planetesimal formation prescription in \eq{dotM_plt} means that streaming instability and planetesimal formation happen when the mid-plane dust-to-gas ratio \hjch{$>\!1$}{exceeds unity}, after which the clumps will exist for a settling time-scale \citep[][]{ShariffCuzzi2015}.
We choose a planetesimal formation efficiency $\zeta = 0.001$ as the default value in our model, which is in line with the choice of other works \citep[e.g.][]{DrazkowskaEtal2016,SchoonenbergEtal2018,StammlerEtal2019}. Yet, a value of $10^{-3}$ seems unphysically low. As we showed in \se{epsilon}, a higher planetesimal formation efficiency implies that the ring will spawn planetesimal much faster and a planetesimal formation efficiency exceeding $\zeta \gtrsim 0.1$ will even consume the ring as shown in JO21. The low $\zeta$ we and other works adopt is partly driven by these reason. A possible solution to this conundrum is to argue that the low $\zeta$ is a manifestation of clump-planetesimal interactions, which tears apart pebble clumps before they gravitationally collapse. Then, although initially planetesimals form on a settling timescale, the existence of large population of planetesimal bodies would quickly suppress further planetesimal formation, and reduce $\zeta$. Future work to test this hypothesis with hydrodynamical simulations is worthwhile.

Our model neglects the gravitational feedback of the planet on the gas, i.e., gap opening for massive planets. A gap will both change the pebble dynamics and slow down the migration of the planets. An upper mass above which our model stops being valid is when the planet strongly perturbs the gas disc (\eq{M_gap}), corresponding to $60\,M_\oplus$ at $74.3$~au in our standard models. \hjch{Such a}{This} mass is never reached in the runs that feature migration. However, \hjrem{the gravitational feedback of planets on the gas disc matters}when there is no migration \hjadd{or when a planet is trapped,} the \hjch{oligarch}{planet} keeps \hjadd{on} growing in situ\hjch{. We discuss the implication of the following evolution in}{, see} \se{insitu}.

Finally, a potential important omission of our model is the lack of gas accretion. Qualitatively, gas accretion is thought to be suppressed when the planet is still accreting \hjch{pebbles}{solids (pebbles and planetesimals)} because gravitational energy released by the accreted solids balances the heat radiated away by the envelope, i.e., the embryos are too hot to accrete gas \citep[e.g.,][]{Rafikov2006,LeeChiang2015,AlibertEtal2018,GuileraEtal2020}. In addition, gas accretion can be suppressed by the the recycling of the atmospheres \citep{MoldenhauerEtal2022} and by the enhanced opacity contributed by the accreted pebbles \citep{OrmelEtal2021}. On the other hand, the polluted atmospheres are characterized by an increased mean molecular weight, which would decrease the critical core mass \citep{HoriIkoma2011,VenturiniEtal2015,BrouwersOrmel2020,BrouwersEtal2021}.
In any case, when the planet moves away from the ring region and pebble accretion subsides, gas accretion is expected to accelerate and the planets formed in our simulations may enter runaway gas accretion. 

\subsection{The “fine-tuned” optical depths of the ring}
One striking feature of the ALMA rings in protoplanetary discs is that their derived peak optical depths are relatively uniform. \citet{DullemondEtal2018} find that the peak optical depths of rings in DSHARP are all around $0.4$, and none of the rings are very optically thick. This result, which is seemingly "fine-tuned", suggests there exists mechanism(s) regulating the upper limit of the rings' optical depths.

Several works have attempted to explain the fine-tuned optical depths. First, the scattering opacity may play a significant role in the interpretation of (sub)millimeter observations \citep[e.g.,][]{KataokaEtal2015,Liu2019}. \citet{ZhuEtal2019} pointed out that dust scattering can reduce the emission from a highly optically thick ring and mimic it as a marginally optically thick ring. Yet, the albedo of dust grains in discs is uncertain \citep[][]{BirnstielEtal2018}. Therefore, in order to answer the question whether or not scattering plays an important role, we need to conduct longer-wavelength observations with upcoming facilities like ngVLA \citep[e.g.,][]{RicciEtal2018}.

It has also been suggested that the fine-tuned optical depth is the consequence of ongoing planetesimal formation \citep{StammlerEtal2019}. As planetesimals form when the mid-plane dust-to-gas ratio reaches unity, the further growth of the ring will be limited, which regulates the optical depth, as we studied in this work. \citet{StammlerEtal2019} studied the case where the ring is formed by a permanent pressure bump. Several hundreds of Earth masses in planetesimals are formed inside the ring, while the further interaction among these planetesimals and with the pebbles in the ring is neglected. We complement the picture by taking pebble accretion, planetesimal dynamics, and migration into account. For the scenario where the inward migration is applicable and a steady state emerges, the saturated ring peak density then explains the fine-tuned ring optical depth naturally. Specifically, for the predicted ring with peak surface density $f_{\rm rs}\Sigma_{\rm tp}$ (\eq{Sig_pp}), the optical depth reads
\begin{equation}
    \tau_{\nu} = \kappa_{\nu}f_{\rm rs}\Sigma_{\rm tp}
\end{equation}
where the opacity $\kappa_{\nu}$ is expressed in terms of particle size $s$, internal density $\rho_{\rm s}$ and absorption coefficient $Q_{\nu,\rm abs}$:
\begin{equation}
    \kappa_{\nu} = \frac{\pi s^2}{m_{\rm s}} Q_{\nu,\rm abs} = \frac{3}{4}\frac{Q_{\nu,\rm abs}}{s\rho_{\rm s}}.
\end{equation}
Since the Stokes number $St = \frac{\pi}{2}\frac{s\rho_{\rm s}}{\Sigma_{\rm g}}$ in the Epstein drag regime, we obtain
\begin{equation}\label{eq:tau_nu}
\begin{aligned}
    \tau_{\nu} &= 
    \frac{1}{2.9} \frac{3\pi}{8} \frac{f_{\rm rs}Q_{\nu,\rm abs}\dot{M}_{\rm net}}{St^{\frac{5}{3}}q_{\rm rp}^{\frac{2}{3}}\Omega_K r_0^2\Sigma_{\rm g}} \\
    & = 0.4
    \left(\frac{Q_{\nu,\rm abs}}{0.05}\right)
    \left(\frac{f_{\rm rs}}{1.8}\right)
    \left(\frac{f_{\rm rp}}{1.4}\right)^{-\frac{2}{3}}
    \left(\frac{f_{\rm mg}}{1}\right)^{\frac{1}{3}}
    \left(\frac{\dot{M}_{\rm net}}{50\,M_\oplus\,\rm Myr^{-1}}\right)^{\frac{2}{3}}
    \\
    &\times
    \left(\frac{\Sigma_{\rm g}}{3.6\,\rm g\,cm^{-2}}\right)^{-\frac{2}{3}}
    \left(\frac{w_{\rm ring}}{3.38\,\rm au}\right)^{-\frac{1}{3}}
    \left(\frac{St}{0.01}\right)^{-\frac{5}{3}}
    \left(\frac{h_{\rm g}}{0.07}\right)^{-\frac{2}{3}}
    \left(\frac{M_\star}{M_\odot}\right)^{-\frac{1}{3}}.
\end{aligned}
\end{equation}
In the Rayleigh regime ($\lambda = c/\nu \gg s$), the absorption coefficient $Q_{\nu,\rm abs} \propto s \propto \mathrm{St}$ \citep[e.g.][]{vandeHulst1957}, which mitigates the St-dependence of the above expression.
Specifically, we find that $Q_{\nu,\rm abs} = 0.05$ for $200\,\mu\mathrm{m}$-sized particles (consistent with our default model) at wavelength of $1.25\rm mm$, based on the Mie opacity calculation provided by \citet{BirnstielEtal2018}. This value agrees with the typical DSHARP ring optical depth.

\subsection{In-situ formation vs factory assembly}
\label{sec:insitu}
Depending on the strength of migration, our model outputs can be classified into two groups: in-situ formation where one planet remains at the ring location and factory assembly where planetary embryos constantly form and migrate inward. \hjadd{In \se{nmg} and \se{real_bump}, we have shown that the case that features a migration trap falls in the in-situ category as the planet cannot escape the ring.} For the in-situ formation case, the biggest planet reached the gap opening mass (\eq{M_gap}) as shown in \fg{stage_nmg}. Such changes in the gas profile will influence the aerodynamics of pebbles around the ring, which we do not model with our simulation. By then, the ring should have already vaporized, due to dominant pebble accretion and the repulsion of the pebbles \hjch{due to}{caused by} the emerging gas gap.

Self-destruction of rings by the formation of a massive planet inside has also been found by other works. \citet{LeeEtal2022} run 2D hydrodynamic simulations investigating the growth of a single planet embryo inside a pebble ring supported by Gaussian profile pressure bump as same as we assume. As they found, the pebble ring will be ingested by the growing planet within ${\sim}1$ Myr (see \citet{CumminsEtal2022} for a similar conclusion). \hjch{During the clean-up stage, asymmetric substructure may form, e.g., the arc substructure in the continuum of Oph--IRS~48 citep{vanderMarelEtal2013i,CumminsEtal2022}.}{This is consistent with our runs without migration (see \fg{stage_nmg} and \fg{stage_d_bump})}.

Therefore, \citet{LeeEtal2022} argue that the rapid evaporation of the ring will shorten its observational window, conflicting the ubiquitous nature of ALMA rings. They state that the formation of planets in ALMA rings must be either a rare or a slow process. \hjadd{This "issue" can be easily solved if the planet formation is inefficient, like in \citet[][]{Morbidelli2020}. By taking a small Stokes number ($3\times10^{-3}$) and finite mass budget inside the ring, the growth of the planetary seed remains slow at large distance \citep[][]{Morbidelli2020}, and therefore the seeds will always coincide with their initial location (but see the discussion in \se{real_bump})}. \hjch{Accounting for migration, we show that a}{If, on the other hand, planets can migrate away from the} ring\hjadd{, our results indicate that rings} can be both long-lived and spawn planets\hjrem{, which reconciles the high frequency of rings and the rapid growth of planets}.

\subsection{\hjch{Implication of the planet formation model}{Exoplanet---Disc ring connection}}
\hjadd{The most frequently observed exoplanets are the close-in Super earth and mini-Neptune \citep[e.g.,][and references therein]{ZhuDong2021}. It has been argued that these compact planet system have their cradle at a more distant location, for example the water iceline \citep[e.g.,][]{OrmelEtal2017,SchoonenbergEtal2019,LiuEtal2020,Coleman2021,IzidoroEtal2021}. Yet, resolving a pebble ring in the inner disc is challenging for ALMA. However, as the planets spawned by the distant ALMA rings in our study are massive, it is conceivable that they will further evolve into giant planets. This raises the hypothesis that the rings seen in ALMA are related to the distant planets found with direct imaging methods}. \hjrem{In this work, we investigated a scenario where planets form in distant rings. Observational signatures of protoplanets are found in several discs.}
\hjch{Here}{Therefore}, we compare our simulation outputs with several \hjadd{observed distant exoplanet} systems.

\subsubsection{\hjch{Planet-ring connection}{ALMA ringed discs that may (not) have planets}}
Since we have modeled our setup on the B74 dust ring in AS~209 --a typical DSHARP ring-- it is justified to ask where its planets are? 
The hunt for detecting planets in DSHARP discs has been attempted through direct imaging \hjrem{of the planets} (e.g., VLT/MUSE, \citet[][]{XieEtal2020}; VLT/NaCo, \citet[][]{JorqueraEtal2021}) and resolving the potential circumplanetary disc \citep[e.g.,][]{AndrewsEtal2021}. Indirectly, by studying the CO kinematic feature, it has been suggested that planets are responsible for nine localized deviations from Keplerian rotation in the kinematic map of eight DSHARP discs \citep[e.g.,][]{PinteEtal2020}. Except for one candidate kink in HD~163296 and \hjadd{a recently identified one in AS~209 \citep[][]{BaeEtal2022}}, all the other eight are located interior to a pebble ring. A standard explanation is that these rings are supported by pressure bumps generated by these hypothesized interior planets. \hjadd{Still, this does not preclude a scenario where the ring has spawned a planet, migrated inwards, and only then generated the velocity kink.}

\hjch{Particularly}{Recently}, it has been proposed that an accreting planet is present in the continuum ring of HD 100546. \citet{CasassusPerez2019} detected a Doppler-flip in the $^{12}\rm CO$~2--1 line centroid, which \citet{CasassusEtal2022} interpreted in the context of a ${\sim}10\,M_\oplus$ accreting planet launching an magnetocentrifugally driven outflow. This can cause a disc eruption, and therefore could lead to a surface disturbance to the Keplerian flow, explaining the flip and localized SO line emission. Such an embedded planet\hjadd{, co-located with the ring,} would be in agreement with our model.

\subsubsection{PDS~70}
With two Jovian mass planets detected in the inner cavity, the transitional disc PDS~70 is currently the only protoplanetary disc with unambiguous evidence of harboring planets \citep{KepplerEtal2018,HaffertEtal2019}. \citet{BaeEtal2019} studied the architecture of the PDS~70 system by hydrodynamical simulation and reproduces the orbital elements of the two planets as well as the ALMA continuum. However, in their work, they initially set the mass of PDS~70~b as $5\,M_{\rm Jup}$ and put it directly at 20~au regardless of the formation history. Considering the 5.4~Myr-old age of PDS~70, the core formation of PDS~70~b, c is inefficient in a smooth disc. For example, if the critical core mass of the PDS~70 protoplanet was $10 M_\oplus$ and the pebble mass flux amounted to $100\, M_\oplus\,\rm Myr^{-1}$, it requires ${\sim}10$~Myr for a Pluto-mass embryo at 20~au to grow \textit{in situ} by pebble accretion in the 3D accretion regime. The growth timescale is even longer for PDS~70~c. 

In addition, the orbits of PDS~70~b, c are consistent with a 2:1 mean motion resonance configuration \citep[e.g.,][]{WangEtal2021i}. \hjch{The appearance of this first-order mean motion resonance hints that they}{If confirmed, the planets} may \hjadd{have} formed in sequence \hjch{in the more outer region and}{even further out to} experience convergent migration \citep{TerquemPapaloizou2007}. \hjch{which will}{This would} require an even longer \hjch{pebble accretion}{planet formation} timescale. \hjch{Therefore,}{To alleviate the timescale problem,} mechanisms to accelerate \hjadd{planet} growth \hjrem{of the protoplanet} are required.

Our model of planet assembly in rings \hjch{has the potential to explain the formation history of}{can potentially be applied to} PDS~70~b and c, i.e., the \hjch{two planetary cores}{planets} are \hjch{formed from}{spawned by} an outer \hjch{first-generation}{pebble} ring, which may still \hjch{appear}{be present} \hjadd{in ALMA imagery} as \hjch{the}{a} continuum peak at ${\sim}74$~au \citep[][]{KepplerEtal2019}. 
Since planet b formed earlier than planet c, it also started gas accretion earlier and transform into the slower type-II migration before planet c \hjadd{caught up}. Together with the fact that PDS~70~c is more massive than PDS~70~b and thus experienced stronger type-I migration torque, it is reasonable to assume that they will undergo convergent migration, and then form a common gap \citep[e.g.,][]{DongFung2017}.

\subsubsection{HR~8799 and other planet systems in debris disc}\label{sec:HR8799}
\hjrem{Apart from planet-forming discs,} Another line of evidence for planet formation in rings comes from the planets detected in debris discs. By statistically comparing the dust mass budget in protoplanetary discs and debris disc, \citet{MichelEtal2021} propose that debris disc are the leftovers of planetesimal-forming rings. This idea is supported by recent theoretical works \citep[e.g.,][]{JiangOrmel2021,NajitaEtal2022}.

\hjrem{Among debris discs,}A number of sources are \hjch{also confirmed with}{seen to harbour} mature planet systems inside the debris disc belt, \hjch{e.g.,}{of which} the four giant planet system HR~8799 \hjadd{is arguably the most iconic}. With years of follow-up direct imaging the orbital configuration of the four giant planets in HR~8799 has been shown to be consistent with a 1:2:4:8 mean motion resonance chain \citep{WangEtal2018}, which, again, hints at the convergent migration of planets. In addition, a clear dust ring is resolved in ALMA (sub)millimeter continuum, whose radial intensity profile peaks around 150 au \citep[e.g.,][]{BoothEtal2016,FaramazEtal2021}. By modeling the radiation transfer in ALMA band 6 and band 7, and reproducing both the Hershel far-infrared map and ALMA continuum, \hjadd{the simulations by} \citet{GeilerEtal2019} infer a total mass of ${\sim}100\,M_\oplus$ of planetesimals up to ${\sim} 100\rm\, km$ \hjrem{by collisional simulations}, \hjadd{a number} which qualitatively matches the mass budget of the remaining planetesimals in our model (\se{plt_belt}). 

In the context of our model, the four giant planet cores \hjadd{would} form from a planetesimal-forming pebble ring \hjadd{situated at} around 150 au. The cores \hjadd{would} migrate inward one by one, evolve into gas giants, and form the Laplace resonance chain. After the gas disc has dissipated, the leftover planetesimals belt from the birth ring \hjch{stays in situ and now constitutes a}{remains to form the} debris disc belt. Detailed modeling of the formation history of PDS~70 and HR~8799 with proper treatment of gas accretion and type-II migration will be studied in a future work. 

\hjrem{In addition,}Observationally resolved debris disc belts are typically much wider than rings in protoplanetary discs \citep[][]{Marino2022}. If rings in protoplanetary discs and debris disc are related, it leaves open the question of how to reconcile their different widths. \citet{MillerEtal2021} proposed that this can be explained by a planetesimal-forming ring migrating inward together with the planet. In JO21, we also showed that the clumpy ring can migrate outward since the pebble supply is upstream. In this work, with fixed ring location , we show that the dynamical interaction between planet and planetesimal can also widen the debris disc ring width (see \fg{n_plt_t}). Debris disc formed in this way should have a sharper inner edge and flatter outer edge (see \se{plt_belt}). Such an asymmetry profile is found in several debris disc systems in the ALMA continuum, e.g., HR~8799 \citep[][]{FaramazEtal2021} and q1~Eri \citep[][]{LovellEtal2021i}.

\section{Conclusions}\label{sec:conclusions}
Rings are the locations where pebbles accumulate to elevate the dust-to-gas ratio, to create conditions conducive to planetesimal formation and further planet assembly. We considered two designs for the rings. One is the clumpy ring (CR) supported by robust aerodynamic backreaction, and the other is the pressure bump (BP) ring which features a Gaussian profile pressure bump. Both designs can reproduce the observations by ALMA in the (sub)millimeter continuum. We have followed the growth of planetesimals formed in these massive pebble ring by N-body simulations. \hjch{Type-I}{Planet} migration
plays a decisive role in the evolution of the ring. Our main findings are\hjrem{ as follows}:

\begin{enumerate}
    \item Pebble rings are ideal locations for planet formation at large orbital distances as the restraining conditions that normally limit pebble accretion no longer apply. The high concentration of pebbles and the low headwind velocity inside the ring render pebble accretion very efficient.
    \item Consequently, our results hardly depend on the initial planetesimal mass. We test that planetesimals whose mass is down to Ceres immediately enjoy effective pebble accretion\hjrem{, while the conclusion can be extended to lower initial mass generally}.
    \item At a distance of 74.3~au and an external mass flux of $100\, M_\oplus\,\rm Myr^{-1}$, a ${\sim}10\,M_\oplus$ planet \hjch{grows from planetesimals formed inside the ring}{emerges} within 0.5~Myr.
    \hjch{Planet formation in rings exhibit similar characteristics,}{This finding is} independent of the mechanism supporting the ring (aerodynamically-supported clumpy ring model or pressure bump-induced). 
    \hjadd{However,} for the CR model the ring mass required to trigger planetesimal formation is relatively modest due to its small scaleheight, while for PB-supported rings this depends on the pebble scaleheight ($\delta_z/\mathrm{St}$) and could be high.
    \item For discs in which inward type-I migration operates across rings, planets migrate away from the pebble ring after reaching a typical mass of ${\sim}20\,M_\oplus$ (\eq{m_rpr}). If external conditions will remain similar, the process will repeat and the ring will spawn a new planet.
    \item The ring's mass is a balance between the consumption by pebble accretion and the external supply. The mass of the ring is typically ${\sim}30\,M_\oplus$ (\eq{M_ring_ss}) at $74.3\,\rm au$, consistent with the mass of annular rings observed in ALMA. 
    \item Planet-forming rings at larger orbital radii are more massive. Yet, the mass of the protoplanet only weakly depends on the locations of the ring. \Eq{Mp_Mring} describes the relation between the observed ring mass and the typical planetary core mass that can form inside the ring.
    \item On the other hand, for disc in which type-I migration is inefficient or rings where planets are trapped (\se{real_bump}), a single big planet will emerge and consume the ring by pebble accretion. In that situation, the window for it to appear as a ring in ALMA imagery could be shorter (${\lesssim}1\,\mathrm{Myr}$).
    \item After the planet migrates away from the ring, its growth by pebble accretion drops rapidly. In the CR design, planets can still grow from the downstream pebble mass flux and it is conceivable that these planets will trigger runaway gas accretion after exiting the ring region.
    \item The remnant planetesimal belts are candidates for the cold debris discs. Scattered or shepherded by the formed planet, the radial profile of the planetesimal belt shows an asymmetry with a steep rise closer at its interior side and a more gradual decline towards the exterior regions -- a feature that may be observable, \se{plt_belt}.
    \item We obtain that the (sub))millimeter optical depth of the planetesimal-forming ring is around ${\sim}0.4$ with minimal dependence on disc and pebble properties (\eq{tau_nu}). This conclusion is in line with observations that \hjch{disc}{ring}s are marginally optically thick \citep[cf.][]{DullemondEtal2018,StammlerEtal2019}. 
\end{enumerate}

The roles of ring and planet recall the classical “chicken or egg” dilemma. According to our model, the massive rings in the protoplanetary disc perform the role of egg-laying chicken. Our model provides an alternative view on the connection between planets and rings than the mainstream model, in which planets induce the formation of the rings.

This leaves open the question of how the first-generation rings form. Due to the great variety in discs properties, it is likely that rings observed in ALMA originate from different mechanisms, e.g., sintering \citep[][]{OkuzumiEtal2016}, secular gravitational instability \citep[][]{TominagaEtal2019}, coagulation front of pebbles \citep[][]{OhashiEtal2021}, anisotropic infall \citep[][]{KuznetsovaEtal2022}. Such a primordial ring can be the birthplace of the first-generation planet, which can later sculpt the disc to possibly generate new rings.

The model where planets form in a pebble ring and then migrate inward has the potential to explain certain systems involving multiple giant planets in young protoplanetary discs and debris discs. Application of this model to the formation history of realistic planetary architecture will be conducted in a follow-up work.

\section*{Acknowledgements}
\hjadd{H.J. and C.W.O. would like to thank the anonymous referee for the constructive report that improved the initial manuscript.} H.J. would like to thank Beibei Liu and Douglas N. C. Lin for their insightful discussions. \hjadd{We also thank Alessandro Morbidelli for useful comments}.
This work has used 
\texttt{Matplotlib} \citep{Hunter2007}, 
\texttt{Numpy} \citep{HarrisEtal2020}, 
\texttt{Scipy} \citep{VirtanenEtal2020} 
software packages.

\section*{Data Availability}
The data underlying this article will be shared on reasonable requests to the corresponding author.
 



\bibliographystyle{mnras}
\bibliography{ads} 




\appendix
\section{Pebble accretion efficiency}\label{app:PAE}
In the main paper we show the pebble accretion efficiency in term of the surface density of pebbles. Our formulas (\eq{RPA}) are converted from the pebble accretion efficiency ($\epsilon_\mathrm{PA} = \dot{M}_{\rm PA}/\dot{M}_{\rm peb}$) in \citet{LiuOrmel2018} and \citet{OrmelLiu2018}. \hjch{Here we show the derivation.}{Substituting the pebble flux $\dot{M}_{\rm peb} = 2\pi r v_\mathrm{dr} \Sigma_{\rm peb}$ with $v_\mathrm{dr} = 2\mathrm{St} \eta v_K$, we define $\mathcal{R} = 4 \pi {\rm St}\eta\epsilon_{\rm PA}$, which is the formula \eq{RPA} in the main text.}

In the 2D and 3D pebble accretion regime, the accretion efficiency are given as
\begin{equation}\label{eq:eps2d}
    \epsilon_{\rm PA, 2D} = \frac{A_2}{\eta}\sqrt{\frac{q_{\rm p}\Delta v}{{\rm St} v_K}} f_{\rm set}
\end{equation}
and
\begin{equation}\label{eq:eps3d}
    \epsilon_{\rm PA, 3D} = A_3 \frac{q_{\rm p}}{\eta h_{\rm peb}} f_{\rm set}^2
\end{equation}
where $q_{\rm p} = m_{\rm p}/M_\star$ is the planet-to-stellar mass ratio, $A_2 = 0.32$ and $A_3 = 0.39$ are fit constant \citep{OrmelLiu2018}. The modulation factor is
\begin{equation}\label{eq:f_set}
    f_{\rm set} = \exp\left(-\frac{1}{2}\frac{\Delta v^2}{v_\ast^2}\right)
\end{equation}
where $v_\ast = (q_{\rm p}/{\rm St})^{\frac{1}{3}}v_K$. Since the high concentration of pebble and the strong dust feedback, the relative velocity between pebble and the planetesimal in the mid-plane of the ring is mainly determined by the Keplerian shear velocity
\begin{equation}\label{eq:v_sh}
    v_{\rm sh} \simeq a_{\rm sh} \left(q_{\rm p} {\rm St}\right)^{\frac{1}{3}} v_K
\end{equation}
where $a_{\rm sh} = 0.52$ is a fit constant from \citet{LiuOrmel2018}. Thus for $\rm St<1$, $f_{\rm set}$ is always closed to unity.

The headwind velocity
\begin{equation}\label{eq:v_hw}
    v_{\rm hw} = \left\{
    \begin{array}{lr}
        \eta_{\rm pb} v_K & \rm PB \\
        0 & \rm CR
    \end{array}
    \right.
\end{equation}
is small since $\eta_{\rm pb}\sim 0$ around the ring peak as shown in \eq{eta_pb} in the PB model, and we ignore the headwind in the CR model since pebbles follow Keplerian velocity in the clumpy ring. Combining $v_{\rm sh}$ and $v_{\rm hw}$, the approach velocity in the circular limit reads
\begin{equation}\label{eq:v_cir}
    v_{\rm cir} = \frac{v_{\rm hw}}{1+a_{\rm cir}q_{\rm p}{\rm St}/\eta^3} + v_{\rm sh}
\end{equation}
where $a_{\rm cir}=5.7$ is another fit constant. In addition, the eccentric velocity
\begin{equation}
    v_{\rm ecc} = a_e e_{\rm p} v_K
\end{equation}
contributes as well, where $a_e = 0.76$ is the numerical fitting constant. Similarly, for the vertical approach velocity we have
\begin{equation}
    v_{\rm inc} = a_i i_{\rm p} v_K
\end{equation}
with fit constant $a_i = 0.68$. Thus the total relative velocity between pebbles and the planetesimal reads
\begin{equation}\label{eq:delv}
    \Delta v = \sqrt{\max(v_{\rm cir}, v_{\rm ecc})^2 + v_{\rm inc}^2}
\end{equation}

\hjrem{By substituting the pebble flux $\dot{M}_{\rm peb} = 2\pi r v_\mathrm{dr} \Sigma_{\rm peb}$ to \eq{eps2d} and \eq{eps3d}, we obtain the formula \eq{RPA} in the main text.}

\section{Trajectory of massive planet}\label{app:mr_tracer}
In the clumpy ring setups, after a planet passes through the inner simulation domain $r_{\rm in} = r_0-5w_{\rm ring}$, it can still accrete from the leaking mass flux $\dot{M}_{\rm leak}$ (\eq{dotM_leak}).

With the default $\delta_{\rm z}=10^{-3}$, pebble accretion away from the ring region is in 3D regime. Therefore, we can analytically calculate the growth trajectory of the planet by solving the pebble accretion rate
\begin{equation}\label{eq:dmdt}
    \frac{{\rm d} m}{{\rm d} t} = \epsilon_{\rm PA}\dot{M}_{\rm leak}
\end{equation}
and the type-I migration speed of planet \citep[e.g.,][]{TanakaEtal2002}
\begin{equation}\label{eq:v_mig}
    \frac{{\rm d} a_{\rm p}}{{\rm d} t} 
    = \dot{L}_p (\frac{{\rm d}L_p}{{\rm d}a_{\rm p}})^{-1} 
    = -\frac{4f_{\rm mg}m\Sigma_{\rm g} \Omega_K a_{\rm p}^3}{M_\star^2 h_{\rm g}^2}.
\end{equation}
Substituting the 3D pebble accretion efficiency \eq{eps3d} into \eq{dmdt}, the differential equation reads
\begin{equation}
    \frac{{\rm d} m}{{\rm d} a_{\rm p}} = -\frac{A_3{\rm St}^{0.5}\dot{M}_{\rm leak}h_{\rm g}}{4\eta\alpha^{0.5}M_\star\Sigma_{\rm g}\Omega_K a_{\rm p}^3} = -\frac{3}{4}\frac{C}{r_0} (\frac{a_{\rm p}}{r_0})^{-0.25}
\end{equation}
where the prefactor $\frac{3}{4}\frac{C}{r_0}$ is independent of $a_{\rm p}$ and $C = 9.1\,M_\oplus$ with the disc parameters in our default run \texttt{cr-default}. The solution to this equation is
\begin{equation}\label{eq:m_r_in}
\begin{aligned}
    m(a_{\rm p}) &= m_{\rm in} + C \times (\frac{r_{\rm in}}{r_0})^{3/4} - C \times (\frac{a_{\rm p}}{r_0})^{3/4} \\
\end{aligned}
\end{equation}
where $m_{\rm in}$ is the mass of the planet when it arrive $r_{\rm in}$. \Eq{m_r_in} is used to plot the trajectory lines in \fg{track}.

\bsp	
\label{lastpage}
\end{document}